\documentclass[aps,pra,twocolumn,groupedaddress,superscriptaddress,longbibliography]{revtex4-1}
\usepackage{color}
\usepackage{amsmath}
\usepackage{amssymb}
\usepackage{graphicx}
\usepackage{epstopdf}
\usepackage[hidelinks,colorlinks=true,linkcolor=blue,citecolor=blue,anchorcolor=black,filecolor=cyan,menucolor=blue,runcolor=cyan,urlcolor=blue]{hyperref}
\usepackage{bm}
\usepackage{color}
\usepackage{txfonts}
\makeatother

\begin{document}

\title{Superfluid Oscillator Circuit with Quantum Current Regulator}
\author{Xue Yang}
\affiliation{Shaanxi Key Laboratory for Theoretical Physics Frontiers, Institute of Modern Physics, Northwest University, Xi'an, 710127, China}
\author{Wenkai Bai}
\affiliation{Shaanxi Key Laboratory for Theoretical Physics Frontiers, Institute of Modern Physics, Northwest University, Xi'an, 710127, China}
\author{Chen Jiao}
\affiliation{Shaanxi Key Laboratory for Theoretical Physics Frontiers, Institute of Modern Physics, Northwest University, Xi'an, 710127, China}
\author{Wu-Ming Liu}
\address{Beijing National Laboratory for Condensed Matter Physics, Institute of Physics, Chinese Academy of Sciences, Beijing, China}
\author{Jun-Hui Zheng}
\email{junhui.zheng@nwu.edu.cn}
\affiliation{Shaanxi Key Laboratory for Theoretical Physics Frontiers, Institute of Modern Physics, Northwest University, Xi'an, 710127, China}
\affiliation{Peng Huanwu Center for Fundamental Theory, Xi'an 710127, China}
\author{Tao Yang}
\email{yangt@nwu.edu.cn}
\affiliation{Shaanxi Key Laboratory for Theoretical Physics Frontiers, Institute of Modern Physics, Northwest University, Xi'an, 710127, China}
\affiliation{Peng Huanwu Center for Fundamental Theory, Xi'an 710127, China}
\date{\today}

\begin{abstract}
We examine the properties of atomic current in a superfluid oscillating circuit consisting of a mesoscopic channel that connects two reservoirs of a Bose-Einstein condensate. We investigate the presence of a critical current in the channel and examine how the amplitude of the oscillations in the number imbalance between the two reservoirs varies with system parameters. In addition to highlighting that the dissipative resistance stems from the formation of vortex pairs, we also illustrate the role of these vortex pairs as a quantum current regulator. The dissipation strength is discrete based on the number imbalance, which corresponds to the emergence of vortex pairs in the system. Our findings indicate that the circuit demonstrates characteristics of both voltage-limiting and current-limiting mechanisms. To model the damping behavior of the atomic superfluid circuit, we develop an equivalent LC oscillator circuit with a quantum current regulator. 
\end{abstract}

\maketitle
\section{\label{sec:level1}INTRODUCTION}

Atomtronics is an emerging interdisciplinary field that focuses on the development of matter-wave circuits utilizing atoms as carriers\,\cite{RevModPhys.94.041001}. The coherence of matter-waves and the many-body effects observed in atomic circuits give rise to novel and exotic behaviors that are not found in electronics or photonics, such as negative differential conductivity\cite{PhysRevLett.115.050601}. Moreover, ultracold atomic gases offer a highly controllable and flexible platform for studying atomic devices, making them the subject of extensive interest for their potential applications in quantum precise measurement\,\cite{PhysRevLett.114.113003, PhysRevLett.84.4749,PhysRevLett.97.200405,PhysRevLett.98.260407}, quantum simulation of various systems\,\cite{PhysRevLett.95.063201,PhysRevA.100.041601,PhysRevA.101.023617,Stadler2012,PhysRevLett.126.055301}, Logic gate\,\cite{PRA.82.013640}, and quantum information processing\,\cite{PhysRevLett.95.173601,Hallwood_2006,Aghamalyan_2016}. To date, several theoretical proposals for atomic devices have been put forth, and a number of them have been successfully demonstrated in experimental settings. These include atomic amplifiers\,\cite{PhysRevLett.91.156403,PhysRevLett.93.140405}, transistors\,\cite{PhysRevLett.101.265302,PhysRevA.75.013608,PhysRevA.90.013616,PhysRevA.104.033311,Caliga_2016}, switches\,\cite{PhysRevLett.102.055702}, batteries\,\cite{Caliga_2017}, memories\,\cite{PhysRevA.85.022318,PhysRevResearch.5.033003,NJP.23.043028}, Josephson junctions\,\cite{PhysRevLett.103.140405,PhysRevA.75.023615,PhysRevA.70.061604,PhysRevResearch.4.033180}, and quantum interference devices\,\cite{PhysRevLett.111.205301,PhysRevA.107.L051303}.

Two-terminal systems have significant potential applications in various mesoscopic atomic optical devices, including quantum metrology, quantum information, and Josephson junctions\,\cite{RevModPhys.90.035005,Haroche,doi:10.1126/science.aaz2342}. Furthermore, these systems are of theoretical importance in elucidating superfluid transport properties such as quantum conductivity\,\cite{PhysRevA.94.023622,PhysRevA.95.013623} and thermoelectric effects\,\cite{doi:10.1126/science.1242308}, as well as constructing general multi-terminal atomic circuits. A simple two-terminal system comprises two reservoirs with a junction. The system exhibits various exotic phenomena depending on the types of reservoir traps, junction structures, and atom properties, each of which is described by different effective models, such as two-mode Rabi oscillations\,\cite{PhysRevLett.120.173601} and inductor-capacitor oscillations\,\cite{PhysRevLett.123.260402}. For a system consisting of two large reservoirs connected by a straight channel (as depicted in Fig.\,\ref{fig1})\,\cite{PhysRevLett.123.260402,PhysRevA.93.063619,PhysRevA.94.053625}, an initial imbalance in the number of atoms between the reservoirs induces a resistive flow during the evolution of the BEC, resulting in subsequent circuit oscillations. This oscillating behavior resembles the adiabatic oscillations observed in superfluid liquid helium transport experiments\,\cite{PhysRev.82.440}. Small initial number imbalances of atoms between the reservoirs induce undamped oscillations of the superfluid flow. An acoustic model can be established to predict the correct frequency of the oscillation by establishing a connection between the kinetic and potential energy contained within sound waves in a superfluid and the electrical energy in an LC circuit\,\cite{PhysRevLett.123.260402}. In the presence of a large population bias initially, the quantum circuit can be analogized as a classical RLC circuit coupled with a Josephson junction\,\cite{PhysRevA.93.063619,PhysRevA.94.023626}. However, for a small population bias, these models fail to capture resistive behaviors and throttling characteristics, and the corresponding theoretical model is considerably less clear.

In this {work}, we investigate a {channel connected} two-terminal system with a small initial population imbalance. Using the Thomas-Fermi approximation, we determine the critical width of the channel for tunneling between two reservoirs. Additionally, we establish the relationship between the oscillatory frequency, amplitude, critical current, and the system's geometry. Furthermore, we develop an equivalent quantum circuit that reproduces the numerical simulation results obtained from the Gross-Pitaevskii equation {(GPE)} of the two-terminal BEC system. Our findings demonstrate a linear increase in the oscillating current amplitude in the channel as the initial population bias increases. Beyond a certain threshold, the emergence of vortex-pairs acts as a quantum current regulator, constraining the oscillating current amplitude to a specific range determined by the energy of a vortex-pair. This discovery suggests the existence of a simple model featuring a quantum current regulator. {The correlation between dissipation and vortex dynamics in our study demonstrates a strong coherence with that found in Josephson-type two-terminal systems.\cite{PhysRevLett.120.025302,PhysRevLett.124.045301,Science.aac9725}.}

\section{Description of superfluid circuit system}\label{sec2}

\begin{figure}[tbp]
\begin{center}
\includegraphics[width=0.95\columnwidth]{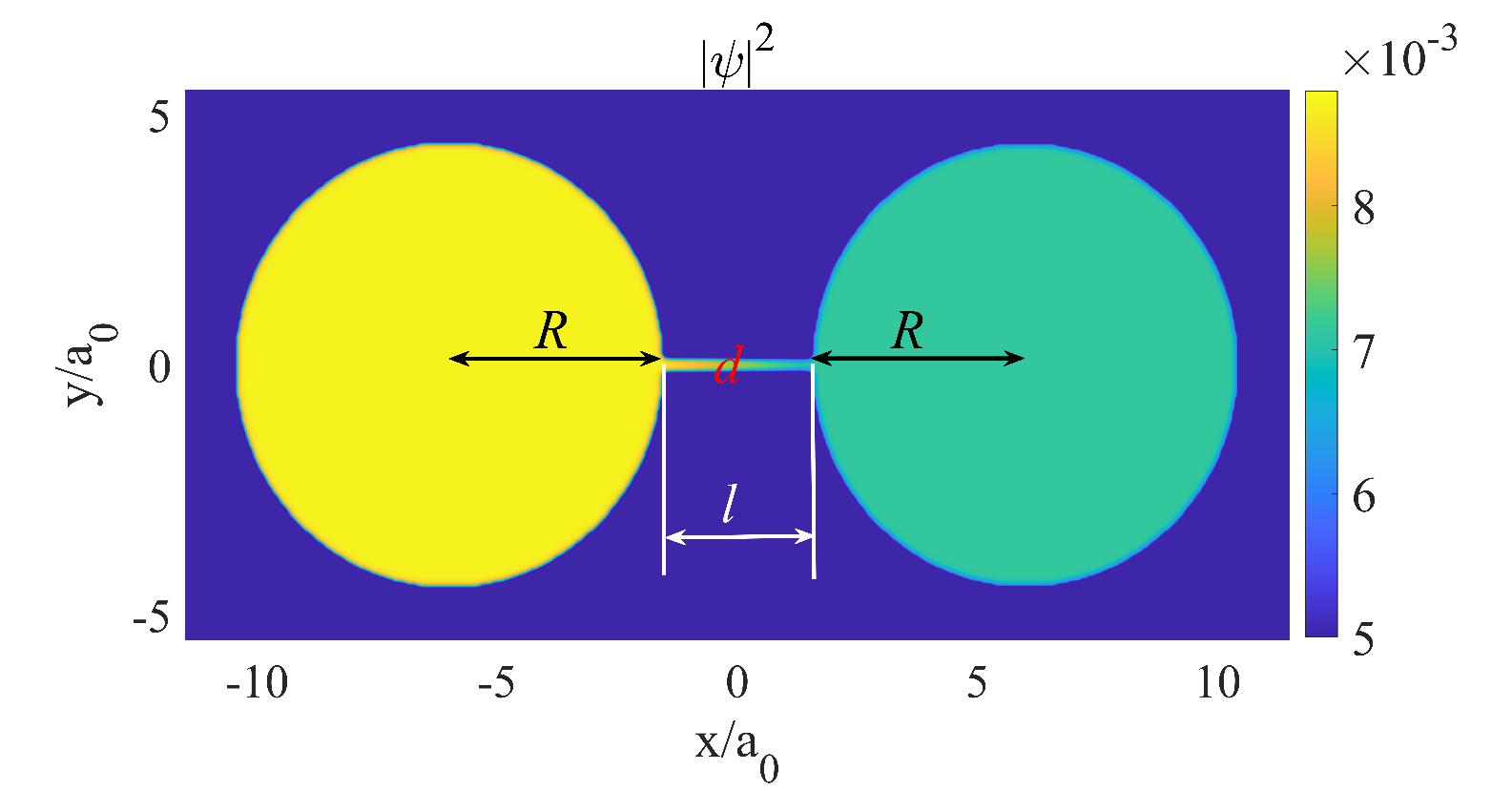}
\caption{The initial density distribution of the BEC in the oscillator circuit consisting of two reservoirs of radius $R=4.5 a_{o}$ and a narrow channel of length $l = 3 a_{o} $ and width $d=0.4 a_{o}$. Here, the potential bias is $ V =0.1  V_c$.}
\label{fig1}
\end{center}
\end{figure}

A superfluid oscillator circuit \cite{PhysRevLett.123.260402, PhysRevA.93.063619} is realized by loading a BEC of ${}^{87}$Rb into a quasi-two-dimensional (quasi-2D) trap {$V_{trap}$} with two reservoirs connected by a narrow channel of length $l$ and width $d$ (see Fig.\,\ref{fig1}). The dynamics of the BEC are governed by the 2D GPE,
\begin{equation}\label{gpe}
i\hbar\frac{\partial \psi}{\partial t}=-\frac{\hbar^2}{2m}\nabla^2\psi+{\left[V_{trap}+\mathcal{V}(r, t)\right]}\psi+g|\psi|^2\psi,
\end{equation}
where $g$ is the effective interatomic interaction strength. In our simulations, $\psi$ is normalized to total number of atoms $N$ and $g = 2.5\times 10^4 \hbar^2/m$.
The potential outside the reservoirs and the channel is a hard-wall potential with a height of $10^5 \hbar\omega_{o}$, where $\omega_{o}=2\pi\times 5 $ Hz is chosen as the reference frequency. In the dynamical evolution, the characteristic time $t_0=1/\omega_{o}$ is chosen as the unit of time.

Inside the trap, the potential is {initially set to be}
\begin{equation}
\mathcal{V}(\bm r, {t=0})=\left\{
\begin{array}{ll}
 0  & ~\text{Left reservoir}\\
 V /2 + x V /l& ~\text{Channel}\\
 V    & ~\text{Right reservoir}
\end{array} \right.,
\end{equation}
 where $V$ refers to the potential bias between two reservoirs. {This potential induces an initial state biased towards population. To drive superfluid flow within the system, $\mathcal{V}(\bm r)$ is maintained at 0 during the dynamics.} The radius of the reservoirs is set to be $R=4.5a_{o}$ with $a_0 = \sqrt{\hbar/m\omega_{o}} \simeq 4.83 \mu m $ being the characteristic length.

The initial state is prepared as the ground state of the BEC system, which can be obtained numerically by solving the GPE in the imaginary time evolution. 
In our numerical calculations, we use a grid size of $800\times 300$ in the spatial dimensions, with spacings $dx = dy =0.05 a_o$. The left reservoir is populated by a higher number of atoms to create the bias of the potential. We define the atom number imbalance {$\eta\doteq{N_L-N_R}$}, where $N_L$ and $N_R$ represent the atom numbers in the left and right reservoirs, respectively. Within the Thomas-Fermi (TF) approximation, when the potential bias exceeds the critical value, given by $V_c = {g}/{\pi R^2} \simeq 393 \hbar \omega_{o}$, the system becomes fully polarized ($\eta =1$). For $ V \leq V_c$, we have {$\eta = {V}/{ V_c}$}. Therefore, the initial number imbalance can be adjusted by linearly changing the potential bias. The numerical result of the BEC density distribution for $V=0.1V_c$ is depicted in Fig.\,\ref{fig1}.

\begin{figure}[tbp]
\begin{center}
\includegraphics[width=0.95\columnwidth]{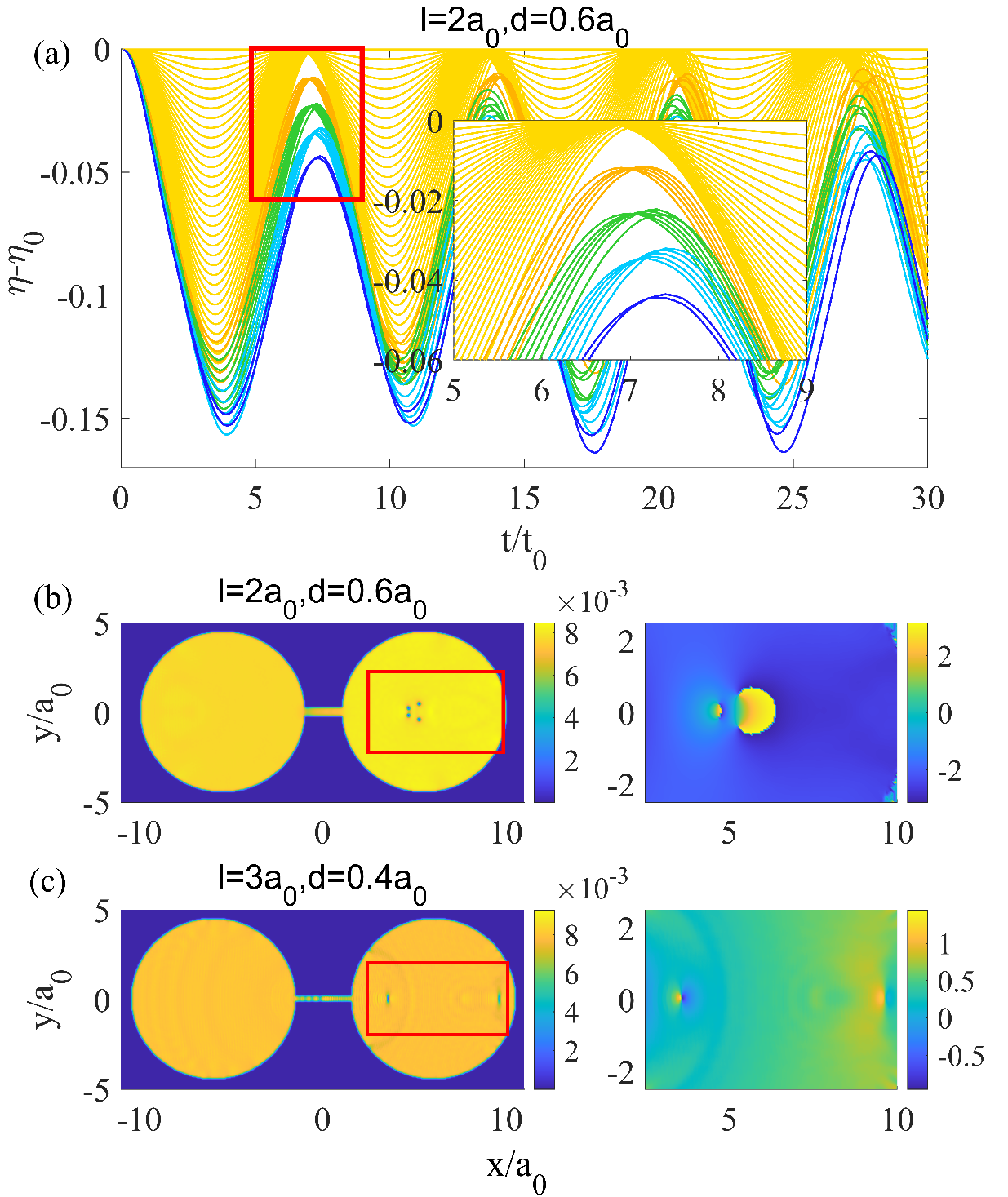}
\caption{(a) The time evolution of the shifted number imbalance, $\eta(t)-\eta(0)$, for 51 different $ V $ uniformly located in the region $ [0, 0.1]  V_c$. The variable $V$ is observed to gradually increase from the yellow lines to the blue lines. The inset is the enlargement of the region in the red box. (b) The density distribution (left subfigure) and the corresponding phase distribution (right subfigure) in the red box region at $t=2.5t_0$ with $V = 0.08 V_c$. The length and width of the channel in (a) and (b) are $l=2a_o$ and $d=0.6a_o$, respectively. (c) The density and the phase distributions at $t=2.5t_0$ with $V = 0.062 V_c$, $l=3a_o$ and $d=0.4a_o$.}
\label{fig2}
\end{center}
\end{figure}

\section{Laws in superfluid oscillator circuit}\label{sec3}

The quench dynamics of the system are investigated by abruptly turning off the potential bias, given as {$\mathcal{V}(\bm r)=0$} for $t>0$. In response, the condensate starts to flow between the left and right reservoirs, causing the number imbalance $\eta(t)$ to change over time. The current in the channel can be expressed as $I={d\eta}/{2 dt}$. Figure\,\ref{fig2}(a) depicts the temporal evolution of the shifted number imbalance, $\eta(t)-\eta(0)$, under various values of $V$ belonging to the range $[0, 0.1] V_c$. {Depending on the magnitude of $V$, the behavior of the system can be categorized as either non-dissipative or dissipative, as indicated by the evolution of the number imbalance.}

In the case where $\eta(0)$ is small, the oscillating number imbalance exhibits a simple cosine behavior without any dissipation. This can be described by the equation $\eta(t) = \eta(0) \cos \omega t$. During this oscillation, the interatomic interaction energy and kinetic energy carried by the flow can transfer into each other. However, when $\eta(0)$ exceeds a certain threshold, the atom number imbalance initially undergoes a decay due to the formation of vortex-pairs near the contact points of the channel and the right reservoir (as depicted in Fig.\,\ref{fig2}(b)). These vortex-pairs are formed because the velocity of the flow in the channel exceeds the critical value $\varv_c$\,\cite{PhysRevA.93.063619}, which leads to energy cascading to small scales and nonlocal kinetic energy dissipation. This dissipation indicates that the flow becomes resistive once the velocity surpasses the critical value. {After the dissipation, the reduced number imbalance recovers and continues to exhibit undamped oscillation.} It is important to note that when the channel width is sufficiently narrow, {dark} solitons rather than vortices are created. These solitons quickly decay into phonons, as shown in Fig.\,\ref{fig2}(c). One can see that compared to vortex excitation, phonon excitation has little effect on the amplitude of oscillating atomic flow. The long-term evolution of $\eta$ is not highly regular, as vortex pairs experience complex movement that induces density wave oscillations (sound waves) within the system \cite{PRA.87.023603,SR.6.29066,FOP.18.62302,RP.22.103828}.

\begin{figure}[tbp]
\begin{center}
\includegraphics[width=0.95\columnwidth]{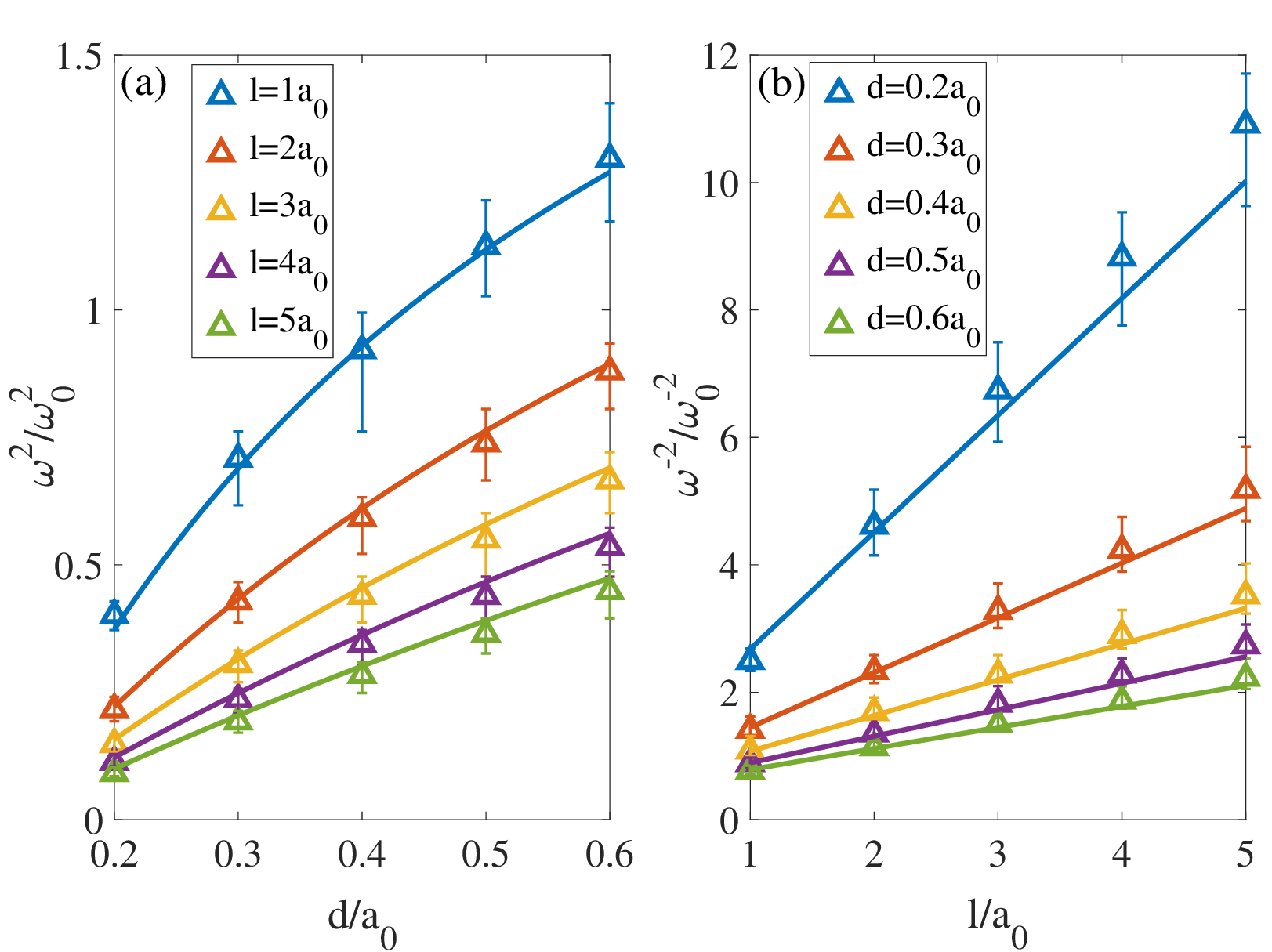}
\caption{The dependence of $\omega^2$ (or $\omega^{-2}$) on two factors of the channel: (a) the width, $d$, and (b) the length, $l$. The triangles accompanied by error bars, represent the statistical average of $\omega$ across different $V$ in the range of $[0, 0.1] V_c$. The error bars indicate the range of $\omega^2$ (or $\omega^{-2}$) resulting from all $V$. Additionally, solid lines depict the fitting results obtained from Eq.\eqref{omega} using $\delta = 2.3d$. }
\label{fig3}
\end{center}
\end{figure}

The evolution of the number imbalance $\eta$ is influenced by various factors such as the initial bias $V$, the channel length $l$, and the width $d$. To establish quantitative relationships between the oscillation of $\eta$ and these system parameters, we fit the undamped portion of the evolution curve $\eta(t)$ by using a sinusoidal function. This allows us to determine the oscillation frequency $\omega$ and the amplitude $A$. Based on our analysis, we conclude that the frequency does not significantly depend on the initial bias and is an intrinsic property of the system. This frequency is linked to the system's geometry, as well as particle properties such as mass and interaction strength. The numerical results depicting the frequencies with changes in the channel dimensions are presented in Fig.\,\ref{fig3}. Similar to the 3D system discussed in Ref.\,\cite{PhysRevLett.123.260402}, we propose that for a 2D system,

\begin{equation}
\omega^2=c^2\left[\frac{(d - d_c)\Theta(d - d_c)}{l+\delta}\left(\frac{1}{S_L}+\frac{1}{S_R}\right)\right],
\label{omega}
\end{equation}
where $c=\sqrt{g{n}/m}$ is the sound speed of the superfluid with number density ${n} = 1/2 \pi R^2$, $S_L=S_R=\pi R^2$ are the areas of the two reservoirs, $\Theta$ is the Heaviside function, and $\delta$ is the end correction for the effective length of the channel. Different from the wide channel case analyzed in Ref.\,\cite{PhysRevLett.123.260402}, the system exhibits no particle current in the channel when its width is below a critical value, $d_c$. We explain this threshold as follows: when the potential $V$ is greater than or equal to $V_c$, the chemical potential $\mu$ approximates as $g / \pi R^2$ in the TF approximation, causing all atoms to populate at the left reservoir. The lowest energy for the transverse standing wave in the channel is $E_0 = {\hbar^2 \pi^2}/{2 m d^2}$. Thus, when $\mu < E_0$, i.e., $d < d_c = {\hbar R \pi^{3/2}}/{(2mg)^{1/2}}$, and the channel is long enough, the atoms are unable to pass through the channel after the bias $V$ is switched off. For the current parameter setting, $d_c$ is approximately equal to $0.112 a_o$. By numerically fitting with Eq.\,\eqref{omega}, the only unknown parameter, $\delta$, is determined to be $2.3d$. The fitting results are in excellent agreement with the numerical results, as illustrated in Fig.\,\ref{fig3}.

\begin{figure}[tbp]
\begin{center}
\includegraphics[width=0.95\columnwidth]{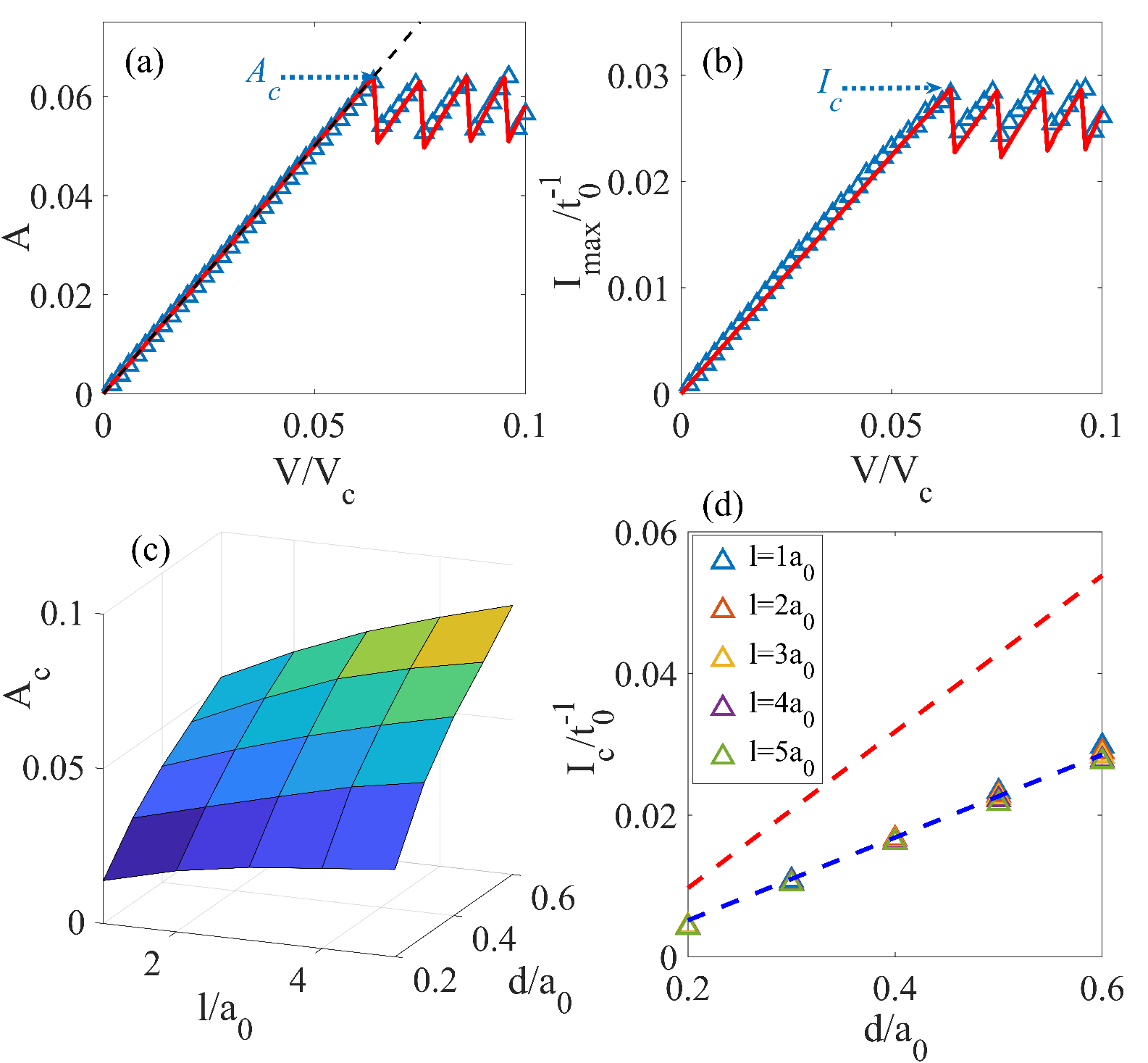}
\caption{(a) The amplitude $A$ and (b) the maximum current $I_{max}$ as functions of $ V/  V_c$ for $l=2 a_o$ and $d = 0.6 a_o$. The black dashed line in (a) is the result of the TF approximation. The red solid lines in (a) and (b) are determined by the equivalent quantum circuit described by Eqs.\,(\ref{lc1}-\ref{lc3}). (c) The critical amplitude $A_c$  and (d) critical current $I_c$ with respect to the geometry of the channel. The blue dashed line in (d) is the fitting result, and the red dashed line is the result given by the Landau critical velocity.}
\label{fig4}
\end{center}
\end{figure}

In Fig.\ref{fig4}(a), the amplitude $A$ of the {dampingless oscillating part} of $\eta(t)$ is plotted against the initial potential bias $V/V_c$ for $l=2a_o$ and $d=0.6a_o$. Initially, when the biases are small, no damping occurs, so we have $A=\eta(0)$, which is equal to $V/V_c$ in the TF approximation. As the bias $V$ increases, the amplitude of the oscillation $A$ also increases. However, when $A$ reaches a critical value, it suddenly decreases and then increases again as the bias increases. This process continues to repeat. Additionally, the instantaneously maximum current passing through the channel, denoted as $I_{max} = \text{Max}[I(t)]$ for {the dampingless oscillating part}, is defined. The changes in $I_{max}$ with the potential bias are shown in Fig.\ref{fig4}(b). Theoretical analysis reveals that $I_{max} = A \omega/2$, which includes characteristics of the amplitude. These characteristics indicate that the circuit is both voltage-limiting and current-limiting. To further analyze, we define the critical amplitude, $A_c$ (the maximum value of $A$ for V in the range $[0, 0.1]V_c$), and the critical current, $I_c$ (the maximum value of $I_{max}$). Figure\,\ref{fig4}(c) and \ref{fig4}(d) demonstrate the plot of these critical values as functions of the channel length and width, respectively. Figure\,\ref{fig4}(d) indicates that the critical value $I_c$ almost does not depend on the length of the channel but linearly depends on the width. This result is reasonable since $I_c$ is proportional to $v_c d_{\text{eff}}$, where $v_c$ is the threshold of superfluid velocity required to generate a vortex-pair and $d_{\text{eff}}= d- d_c$ is the effective width of the channel. Based on Eq.\eqref{omega} and the fact that $A_c = 2I_c/\omega$, it follows that $A_c$ is proportional to $ (d_{\text{eff}} l_{\text{eff}})^{1/2}$, where $l_{\text{eff}}= l + \delta$ is the effective length of the channel. This explains the increase of critical value $A_c$ with $d$ and $l$, as shown in Fig.\,4(c). In the following, we provide a quantitative explanation for the critical value $I_c$.

From the Bogoliubov spectrum of the GPE\,\eqref{gpe}, we can determine that the Landau critical velocity, which is equivalent to the phonon speed, can be expressed as $v_{L} = \text{Min}[E_k/\hbar k] =\sqrt{gn/m}$, where $E_k=\sqrt{\epsilon_k^2+2\epsilon_k g {n}}$ and $\epsilon_k = k^2/2m$ is the kinetic energy of free atoms. Hence, the Landau critical current in the channel can be represented as $I_{L} = v_L d_{\text{eff}} n$. In Fig.\,\ref{fig4}(d), the Landau critical current is illustrated with a red dashed line, while the critical current ($I_c$) obtained from simulations is presented with a blue dashed line, revealing a discrepancy where $I_c=0.53I_L$. This deviation arises due to the fact that the critical velocity ($v_c$) required to generate vortex-pairs is proportional to the Landau critical velocity, {with a ratio of 0.42 for a cylindrical moving obstacle \cite{PhysRevLett.69.1644,PhysRevLett.83.2502}}.

\begin{figure}[tbp]
\begin{center}
\includegraphics[width=0.95\columnwidth]{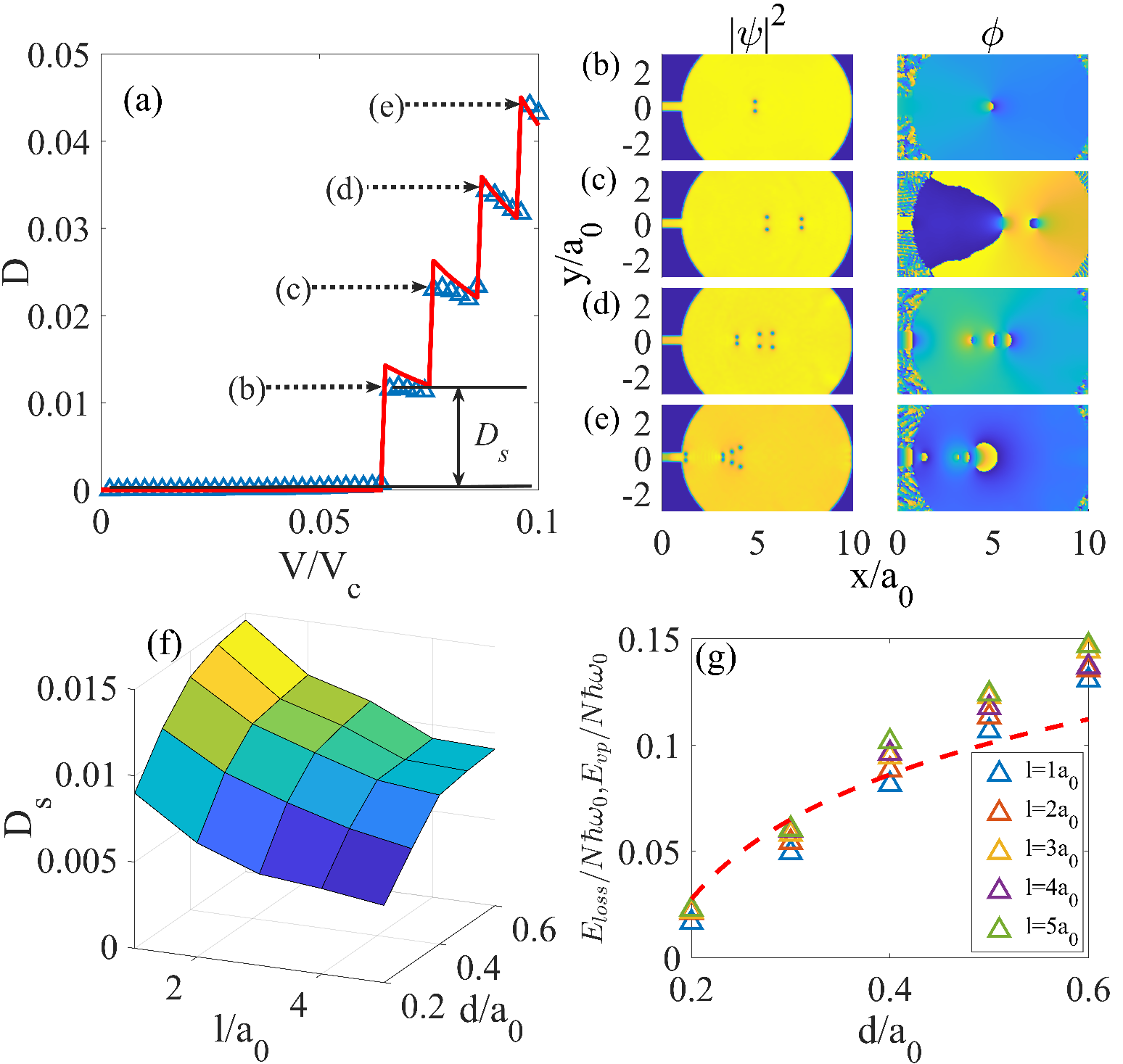}
\caption{(a) The dissipation strength $D = \eta_0 - A$ as a function of the potential bias $V$ for $l=2a_o$ and $d=0.6a_o$. The red solid line is the result of the equivalent quantum circuit described by Eqs.(\ref{lc1}-\ref{lc3}). (b-e) Vortices created for the biases marked in (a). (f) The dissipation step $D_s$ as a function of $l$ and $d$. (g) The energy loss (the blue dashed line) and the energy of a single vortex-pair (triangles) as a function of $d$ for different $l$.}
\label{fig5}
\end{center}
\end{figure}

The damping behavior in the superfluid circuit can be represented by calculating the dissipation strength, defined as $D = \eta(0) - A$. Figure\,\ref{fig5}(a) illustrates that for small biases, the strength $D$ remains at 0. However, it increases abruptly when the bias voltage $V/V_c$ reaches a series of discrete values. Each of these values corresponds to the creation of a new vortex-pair, as depicted in Figs.\,\ref{fig5}(b)-\ref{fig5}(e). Additionally, we introduce the concept of one-step jumping $D_s$ of the dissipation strength, indicated in Fig.\,\ref{fig5}(a). It is worth noting that as the bias voltage increases, the magnitudes of subsequent step jumps become slightly smaller compared to the first step jump due to the interactions between vortex-pairs. The relationship between the first step jumping $D_s$ and the geometric parameters of the channel ($l$ and $d$) is presented in Fig.\,\ref{fig5}(f), revealing that $D_s$ increases with $d$ but decreases with $l$.

The dissipation phenomenon occurs when a portion of the interaction energy in the initial state is utilized in the formation of vortices. By employing the TF approximation, we can estimate the loss of interaction energy during a one-step jump. Assuming an initial bias $\eta(0)$, the total energy of the system immediately after quenching can be approximated as $E=\frac{1}{2}g N ( N_L^2 + N_R^2)/{\pi R^2}$, where {$N=N_L+N_R$}. Simultaneously, within the TF approximation, we have the equation $g( N_L - N_R)/{\pi R^2} = V = \eta(0) V_c$. By considering the fact that $N_L + N_R \approx 1$, and for small $\eta(0)$, we can obtain the initial energy of the system
\begin{equation} \label{energy}
E=\frac{g N [1+ \eta(0)^2]}{4\pi R^2}.
\end{equation}
After the one-step jumping, the bias cannot be recovered any more, i.e., $\eta(0) \rightarrow [\eta(0) -D_s]$. From Eq.\eqref{energy}, the interaction energy loss is about
\begin{equation}\label{eloss}
E_{\text{loss}}=\frac{g N [2 \eta(0) D_s - D_s^2]}{4\pi R^2}.
\end{equation}
The energy of a single vortex-pair is given by the expression $E_{vp}=({2\pi \hbar^2 N n}/{m})\ln({d_{\text{eff}}}/{\xi})$, where $\xi=\hbar/\sqrt{2g n m}$ represents the healing length\,\cite{Feynman}. Figure\,\ref{fig5}(g) provides a comparison between $E_{vp}/N$ and $E_{\text{loss}}/N$ for different widths $d$. The results indicate that the energy of the vortex-pair is approximately equivalent to the loss of interaction energy, thus confirming that the generation of vortex excitations effectively dissipates the number imbalance between the two reservoirs.

\begin{figure}[tbp]
  \centering
  \includegraphics[width=3 in]{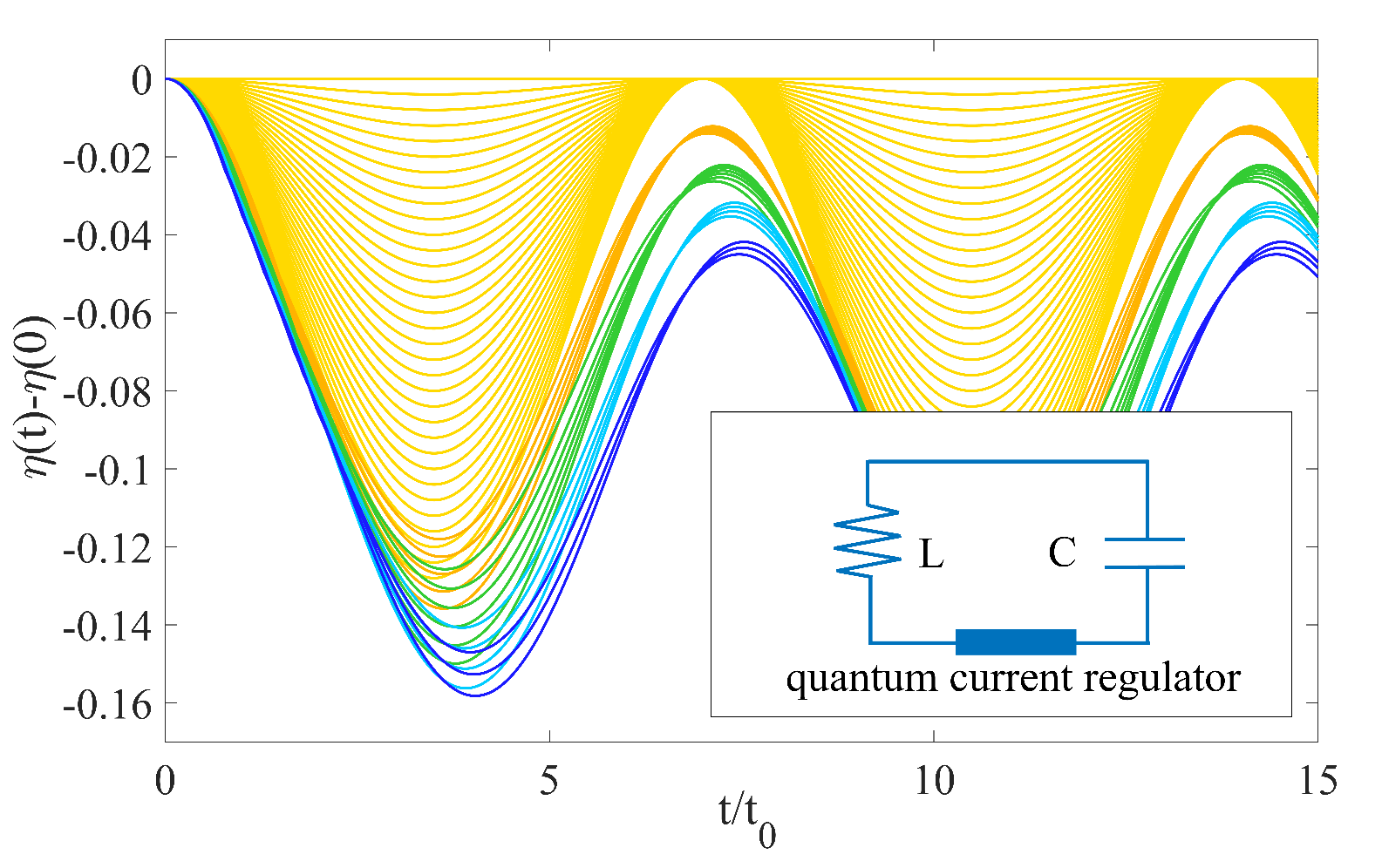}\\
  \caption{The time evolution of $\eta(t)-\eta(0)$ for the equivalent quantum circuit (inset), where $\eta(t) =  2 Q(t)$ with $Q(t)$ being the charges stored in the capacitance. This simulation corresponds to the superfluid circuit of $l=2a_o$ and $d=0.6a_o$ with $\eta(0) \in [0, 0.1]$ as shown in Fig.\,\ref{fig2}. In the simulation, the parameters $\omega=0.8984\omega_0$, $D_s=0.0145$, and $I_c=0.0289\omega_0$ were used. These values correspond to approximately $95\%$ and $141\%$ of the theoretical estimation values for $\omega^{th}=0.9457\omega_0$ (obtained from Eq. (1)) and $D_s^{th}=0.0103$ (derived from $E_\text{loss} = E_{vp}$), respectively. The results for $A$, $I_{{max}}$, and $D$ are shown in Figs.\,\ref{fig4}(a), \ref{fig4}(b), and \ref{fig5}(a) respectively.}\label{fig6}
\end{figure}

\section{Equivalent quantum circuit}\label{sec4}

We have observed that the dissipation $D$ exhibits a step-like behavior in the region $\eta(0) \in [0,0.1]$, corresponding to the initial potential bias or atom number imbalance. The discontinuity in the dissipation is directly linked to the creation of vortices within the system. Each step increase in dissipation signifies the formation of an additional vortex pair. This behavior results in the circuit acting as both a voltage-limiting and current-limiting device. Consequently, the classical RLC circuit analogy is inadequate in this scenario. Instead, an additional quantum current regulator is required to control the maximum current by manipulating quanta. The equivalent quantum circuit is depicted in Fig.\,\ref{fig6}. The current satisfies
\begin{eqnarray}
\dot Q(t) &=& I(t), \label{lc1} \\
\dot I(t) &=& - \omega^2 Q(t), \label{lc2}
\end{eqnarray}
for $|I(t)|<I_c$, where $\omega^2 = 1/LC$ can be determined by Eq.\eqref{omega}, and $Q(t)$ is the instantaneous charge stored in the capacitor. When $|I(t)|$ reaches $I_c$, it suffers a sudden suppress,
\begin{equation}\label{lc3}
  I(t_+) = \text{Sign}[I(t)] (|I(t)|-\Delta I),
\end{equation}
where $t_+ = t + 0^+$, and $\Delta I = D_s \omega/2$ is a quantum, representing the creation of a vortex-pair. $D_s$ can be estimated by setting $E_\text{loss} = E_{vp}$. The mapping between $\eta(t)$ and $Q(t)$ is $\eta(t) = 2 Q(t)$, since $I(t)=d\eta(t)/2dt$. As shown in Fig.\,\ref{fig6}, the equivalent quantum circuit well replays the results given by the original superfluid circuit.

\section{Conclusions}\label{sec5}

In conclusion, we have investigated the behavior of a superfluid oscillating circuit comprised of a mesoscopic channel connecting two large BEC reservoirs. Our study revealed the presence of a critical current that is proportionate to the effective channel width and independent of its length. Additionally, we determined the correlation between the amplitude of oscillation and the system parameters. We demonstrated that the emergence of vortex-pairs serves as a quantum current regulator, leading to intriguing phenomena in this highly non-linear system. Initially, a small {population imbalance $\eta$} resulted in a linear relationship with the induced current's amplitude. However, surpassing a threshold in bias excess constrained the amplitude to the interval of $[I_c - \Delta I, I_c]$. Furthermore, we constructed an equivalent LC oscillator circuit with a quantum current regulator, establishing a comprehensive link between the parameters of the quantum LC circuit and the original superfluid circuit. It is important to note that this correspondence between the two circuits is limited to small initial biases. Interestingly, the complete suppression of oscillation amplitude observed at large $\eta(0) \approx 0.6$ in Ref. \cite{PhysRevLett.123.260402} cannot be accounted for by the current equivalent circuit. In this particular system, the damping of current stems from the creation of vortex-pairs, as opposed to heat dissipation. The energy loss is stored within the vortex-pairs, which could potentially function as a quantum battery if the stored energy can be efficiently released.

\begin{acknowledgments}

We acknowledge Dr. Lijia Jiang for her suggestions. This work is supported by the National Natural Science Foundation of China under grants Nos. 12175180, 11934015, 12247103, 12247186 and 12234012, the National Key R$\&$D Program of China under grants Nos. 2021YFA1400900, 2021YFA0718300 and 2021YFA1402100, the Major Basic Research Program of Natural Science of Shaanxi Province under grants Nos. 2017KCT-12 and 2017ZDJC-32, Shaanxi Fundamental Science Research Project for Mathematics and Physics under grant Nos.\,22JSZ005 and 22JSQ041, the Scientific Research Program Funded by Education
Department of Shaanxi Provincial Government under grant No.
22JK0581, and the Natural Science Basic Research Program of Shaanxi
under grant No. 2023-JC-QN-0054. This research is also supported by The Double First-class University Construction Project of Northwest University.
\end{acknowledgments}


\begin{thebibliography}{55}%
	\makeatletter
	\providecommand \@ifxundefined [1]{%
		\@ifx{#1\undefined}
	}%
	\providecommand \@ifnum [1]{%
		\ifnum #1\expandafter \@firstoftwo
		\else \expandafter \@secondoftwo
		\fi
	}%
	\providecommand \@ifx [1]{%
		\ifx #1\expandafter \@firstoftwo
		\else \expandafter \@secondoftwo
		\fi
	}%
	\providecommand \natexlab [1]{#1}%
	\providecommand \enquote  [1]{``#1''}%
	\providecommand \bibnamefont  [1]{#1}%
	\providecommand \bibfnamefont [1]{#1}%
	\providecommand \citenamefont [1]{#1}%
	\providecommand \href@noop [0]{\@secondoftwo}%
	\providecommand \href [0]{\begingroup \@sanitize@url \@href}%
	\providecommand \@href[1]{\@@startlink{#1}\@@href}%
	\providecommand \@@href[1]{\endgroup#1\@@endlink}%
	\providecommand \@sanitize@url [0]{\catcode `\\12\catcode `\$12\catcode
		`\&12\catcode `\#12\catcode `\^12\catcode `\_12\catcode `\%12\relax}%
	\providecommand \@@startlink[1]{}%
	\providecommand \@@endlink[0]{}%
	\providecommand \url  [0]{\begingroup\@sanitize@url \@url }%
	\providecommand \@url [1]{\endgroup\@href {#1}{\urlprefix }}%
	\providecommand \urlprefix  [0]{URL }%
	\providecommand \Eprint [0]{\href }%
	\providecommand \doibase [0]{http://dx.doi.org/}%
	\providecommand \selectlanguage [0]{\@gobble}%
	\providecommand \bibinfo  [0]{\@secondoftwo}%
	\providecommand \bibfield  [0]{\@secondoftwo}%
	\providecommand \translation [1]{[#1]}%
	\providecommand \BibitemOpen [0]{}%
	\providecommand \bibitemStop [0]{}%
	\providecommand \bibitemNoStop [0]{.\EOS\space}%
	\providecommand \EOS [0]{\spacefactor3000\relax}%
	\providecommand \BibitemShut  [1]{\csname bibitem#1\endcsname}%
	\let\auto@bib@innerbib\@empty
	\bibitem [{\citenamefont {Amico}\ \emph {et~al.}(2022)\citenamefont {Amico},
		\citenamefont {Anderson}, \citenamefont {Boshier}, \citenamefont {Brantut},
		\citenamefont {Kwek}, \citenamefont {Minguzzi},\ and\ \citenamefont {von
			Klitzing}}]{RevModPhys.94.041001}%
	\BibitemOpen
	\bibfield  {author} {\bibinfo {author} {\bibfnamefont {Luigi}\ \bibnamefont
			{Amico}}, \bibinfo {author} {\bibfnamefont {Dana}\ \bibnamefont {Anderson}},
		\bibinfo {author} {\bibfnamefont {Malcolm}\ \bibnamefont {Boshier}}, \bibinfo
		{author} {\bibfnamefont {Jean-Philippe}\ \bibnamefont {Brantut}}, \bibinfo
		{author} {\bibfnamefont {Leong-Chuan}\ \bibnamefont {Kwek}}, \bibinfo
		{author} {\bibfnamefont {Anna}\ \bibnamefont {Minguzzi}}, \ and\ \bibinfo
		{author} {\bibfnamefont {Wolf}\ \bibnamefont {von Klitzing}},\ }\bibfield
	{title} {\enquote {\bibinfo {title} {Colloquium: Atomtronic circuits: From
				many-body physics to quantum technologies},}\ }\href {\doibase
		10.1103/RevModPhys.94.041001} {\bibfield  {journal} {\bibinfo  {journal}
			{Rev. Mod. Phys.}\ }\textbf {\bibinfo {volume} {94}},\ \bibinfo {pages}
		{041001} (\bibinfo {year} {2022})}\BibitemShut {NoStop}%
	\bibitem [{\citenamefont {Labouvie}\ \emph {et~al.}(2015)\citenamefont
		{Labouvie}, \citenamefont {Santra}, \citenamefont {Heun}, \citenamefont
		{Wimberger},\ and\ \citenamefont {Ott}}]{PhysRevLett.115.050601}%
	\BibitemOpen
	\bibfield  {author} {\bibinfo {author} {\bibfnamefont {Ralf}\ \bibnamefont
			{Labouvie}}, \bibinfo {author} {\bibfnamefont {Bodhaditya}\ \bibnamefont
			{Santra}}, \bibinfo {author} {\bibfnamefont {Simon}\ \bibnamefont {Heun}},
		\bibinfo {author} {\bibfnamefont {Sandro}\ \bibnamefont {Wimberger}}, \ and\
		\bibinfo {author} {\bibfnamefont {Herwig}\ \bibnamefont {Ott}},\ }\bibfield
	{title} {\enquote {\bibinfo {title} {Negative differential conductivity in an
				interacting quantum gas},}\ }\href {\doibase 10.1103/PhysRevLett.115.050601}
	{\bibfield  {journal} {\bibinfo  {journal} {Phys. Rev. Lett.}\ }\textbf
		{\bibinfo {volume} {115}},\ \bibinfo {pages} {050601} (\bibinfo {year}
		{2015})}\BibitemShut {NoStop}%
	\bibitem [{\citenamefont {Weiss}\ \emph {et~al.}(2015)\citenamefont {Weiss},
		\citenamefont {Knufinke}, \citenamefont {Bernon}, \citenamefont {Bothner},
		\citenamefont {S\'ark\'any}, \citenamefont {Zimmermann}, \citenamefont
		{Kleiner}, \citenamefont {Koelle}, \citenamefont {Fort\'agh},\ and\
		\citenamefont {Hattermann}}]{PhysRevLett.114.113003}%
	\BibitemOpen
	\bibfield  {author} {\bibinfo {author} {\bibfnamefont {P.}~\bibnamefont
			{Weiss}}, \bibinfo {author} {\bibfnamefont {M.}~\bibnamefont {Knufinke}},
		\bibinfo {author} {\bibfnamefont {S.}~\bibnamefont {Bernon}}, \bibinfo
		{author} {\bibfnamefont {D.}~\bibnamefont {Bothner}}, \bibinfo {author}
		{\bibfnamefont {L.}~\bibnamefont {S\'ark\'any}}, \bibinfo {author}
		{\bibfnamefont {C.}~\bibnamefont {Zimmermann}}, \bibinfo {author}
		{\bibfnamefont {R.}~\bibnamefont {Kleiner}}, \bibinfo {author} {\bibfnamefont
			{D.}~\bibnamefont {Koelle}}, \bibinfo {author} {\bibfnamefont
			{J.}~\bibnamefont {Fort\'agh}}, \ and\ \bibinfo {author} {\bibfnamefont
			{H.}~\bibnamefont {Hattermann}},\ }\bibfield  {title} {\enquote {\bibinfo
			{title} {Sensitivity of ultracold atoms to quantized flux in a
				superconducting ring},}\ }\href {\doibase 10.1103/PhysRevLett.114.113003}
	{\bibfield  {journal} {\bibinfo  {journal} {Phys. Rev. Lett.}\ }\textbf
		{\bibinfo {volume} {114}},\ \bibinfo {pages} {113003} (\bibinfo {year}
		{2015})}\BibitemShut {NoStop}%
	\bibitem [{\citenamefont {Folman}\ \emph {et~al.}(2000)\citenamefont {Folman},
		\citenamefont {Kr\"uger}, \citenamefont {Cassettari}, \citenamefont {Hessmo},
		\citenamefont {Maier},\ and\ \citenamefont
		{Schmiedmayer}}]{PhysRevLett.84.4749}%
	\BibitemOpen
	\bibfield  {author} {\bibinfo {author} {\bibfnamefont {Ron}\ \bibnamefont
			{Folman}}, \bibinfo {author} {\bibfnamefont {Peter}\ \bibnamefont
			{Kr\"uger}}, \bibinfo {author} {\bibfnamefont {Donatella}\ \bibnamefont
			{Cassettari}}, \bibinfo {author} {\bibfnamefont {Bj\"orn}\ \bibnamefont
			{Hessmo}}, \bibinfo {author} {\bibfnamefont {Thomas}\ \bibnamefont {Maier}},
		\ and\ \bibinfo {author} {\bibfnamefont {J\"org}\ \bibnamefont
			{Schmiedmayer}},\ }\bibfield  {title} {\enquote {\bibinfo {title}
			{Controlling cold atoms using nanofabricated surfaces: Atom chips},}\ }\href
	{\doibase 10.1103/PhysRevLett.84.4749} {\bibfield  {journal} {\bibinfo
			{journal} {Phys. Rev. Lett.}\ }\textbf {\bibinfo {volume} {84}},\ \bibinfo
		{pages} {4749--4752} (\bibinfo {year} {2000})}\BibitemShut {NoStop}%
	\bibitem [{\citenamefont {Nirrengarten}\ \emph {et~al.}(2006)\citenamefont
		{Nirrengarten}, \citenamefont {Qarry}, \citenamefont {Roux}, \citenamefont
		{Emmert}, \citenamefont {Nogues}, \citenamefont {Brune}, \citenamefont
		{Raimond},\ and\ \citenamefont {Haroche}}]{PhysRevLett.97.200405}%
	\BibitemOpen
	\bibfield  {author} {\bibinfo {author} {\bibfnamefont {T.}~\bibnamefont
			{Nirrengarten}}, \bibinfo {author} {\bibfnamefont {A.}~\bibnamefont {Qarry}},
		\bibinfo {author} {\bibfnamefont {C.}~\bibnamefont {Roux}}, \bibinfo {author}
		{\bibfnamefont {A.}~\bibnamefont {Emmert}}, \bibinfo {author} {\bibfnamefont
			{G.}~\bibnamefont {Nogues}}, \bibinfo {author} {\bibfnamefont
			{M.}~\bibnamefont {Brune}}, \bibinfo {author} {\bibfnamefont {J.-M.}\
			\bibnamefont {Raimond}}, \ and\ \bibinfo {author} {\bibfnamefont
			{S.}~\bibnamefont {Haroche}},\ }\bibfield  {title} {\enquote {\bibinfo
			{title} {Realization of a superconducting atom chip},}\ }\href {\doibase
		10.1103/PhysRevLett.97.200405} {\bibfield  {journal} {\bibinfo  {journal}
			{Phys. Rev. Lett.}\ }\textbf {\bibinfo {volume} {97}},\ \bibinfo {pages}
		{200405} (\bibinfo {year} {2006})}\BibitemShut {NoStop}%
	\bibitem [{\citenamefont {Mukai}\ \emph {et~al.}(2007)\citenamefont {Mukai},
		\citenamefont {Hufnagel}, \citenamefont {Kasper}, \citenamefont {Meno},
		\citenamefont {Tsukada}, \citenamefont {Semba},\ and\ \citenamefont
		{Shimizu}}]{PhysRevLett.98.260407}%
	\BibitemOpen
	\bibfield  {author} {\bibinfo {author} {\bibfnamefont {T.}~\bibnamefont
			{Mukai}}, \bibinfo {author} {\bibfnamefont {C.}~\bibnamefont {Hufnagel}},
		\bibinfo {author} {\bibfnamefont {A.}~\bibnamefont {Kasper}}, \bibinfo
		{author} {\bibfnamefont {T.}~\bibnamefont {Meno}}, \bibinfo {author}
		{\bibfnamefont {A.}~\bibnamefont {Tsukada}}, \bibinfo {author} {\bibfnamefont
			{K.}~\bibnamefont {Semba}}, \ and\ \bibinfo {author} {\bibfnamefont
			{F.}~\bibnamefont {Shimizu}},\ }\bibfield  {title} {\enquote {\bibinfo
			{title} {Persistent supercurrent atom chip},}\ }\href {\doibase
		10.1103/PhysRevLett.98.260407} {\bibfield  {journal} {\bibinfo  {journal}
			{Phys. Rev. Lett.}\ }\textbf {\bibinfo {volume} {98}},\ \bibinfo {pages}
		{260407} (\bibinfo {year} {2007})}\BibitemShut {NoStop}%
	\bibitem [{\citenamefont {Amico}\ \emph {et~al.}(2005)\citenamefont {Amico},
		\citenamefont {Osterloh},\ and\ \citenamefont
		{Cataliotti}}]{PhysRevLett.95.063201}%
	\BibitemOpen
	\bibfield  {author} {\bibinfo {author} {\bibfnamefont {Luigi}\ \bibnamefont
			{Amico}}, \bibinfo {author} {\bibfnamefont {Andreas}\ \bibnamefont
			{Osterloh}}, \ and\ \bibinfo {author} {\bibfnamefont {Francesco}\
			\bibnamefont {Cataliotti}},\ }\bibfield  {title} {\enquote {\bibinfo {title}
			{Quantum many particle systems in ring-shaped optical lattices},}\ }\href
	{\doibase 10.1103/PhysRevLett.95.063201} {\bibfield  {journal} {\bibinfo
			{journal} {Phys. Rev. Lett.}\ }\textbf {\bibinfo {volume} {95}},\ \bibinfo
		{pages} {063201} (\bibinfo {year} {2005})}\BibitemShut {NoStop}%
	\bibitem [{\citenamefont {Haug}\ \emph {et~al.}(2019)\citenamefont {Haug},
		\citenamefont {Heimonen}, \citenamefont {Dumke}, \citenamefont {Kwek},\ and\
		\citenamefont {Amico}}]{PhysRevA.100.041601}%
	\BibitemOpen
	\bibfield  {author} {\bibinfo {author} {\bibfnamefont {Tobias}\ \bibnamefont
			{Haug}}, \bibinfo {author} {\bibfnamefont {Hermanni}\ \bibnamefont
			{Heimonen}}, \bibinfo {author} {\bibfnamefont {Rainer}\ \bibnamefont
			{Dumke}}, \bibinfo {author} {\bibfnamefont {Leong-Chuan}\ \bibnamefont
			{Kwek}}, \ and\ \bibinfo {author} {\bibfnamefont {Luigi}\ \bibnamefont
			{Amico}},\ }\bibfield  {title} {\enquote {\bibinfo {title} {Aharonov-bohm
				effect in mesoscopic bose-einstein condensates},}\ }\href {\doibase
		10.1103/PhysRevA.100.041601} {\bibfield  {journal} {\bibinfo  {journal}
			{Phys. Rev. A}\ }\textbf {\bibinfo {volume} {100}},\ \bibinfo {pages}
		{041601} (\bibinfo {year} {2019})}\BibitemShut {NoStop}%
	\bibitem [{\citenamefont {Kohn}\ \emph {et~al.}(2020)\citenamefont {Kohn},
		\citenamefont {Silvi}, \citenamefont {Gerster}, \citenamefont {Keck},
		\citenamefont {Fazio}, \citenamefont {Santoro},\ and\ \citenamefont
		{Montangero}}]{PhysRevA.101.023617}%
	\BibitemOpen
	\bibfield  {author} {\bibinfo {author} {\bibfnamefont {L.}~\bibnamefont
			{Kohn}}, \bibinfo {author} {\bibfnamefont {P.}~\bibnamefont {Silvi}},
		\bibinfo {author} {\bibfnamefont {M.}~\bibnamefont {Gerster}}, \bibinfo
		{author} {\bibfnamefont {M.}~\bibnamefont {Keck}}, \bibinfo {author}
		{\bibfnamefont {R.}~\bibnamefont {Fazio}}, \bibinfo {author} {\bibfnamefont
			{G.~E.}\ \bibnamefont {Santoro}}, \ and\ \bibinfo {author} {\bibfnamefont
			{S.}~\bibnamefont {Montangero}},\ }\bibfield  {title} {\enquote {\bibinfo
			{title} {Superfluid-to-mott transition in a bose-hubbard ring: Persistent
				currents and defect formation},}\ }\href {\doibase
		10.1103/PhysRevA.101.023617} {\bibfield  {journal} {\bibinfo  {journal}
			{Phys. Rev. A}\ }\textbf {\bibinfo {volume} {101}},\ \bibinfo {pages}
		{023617} (\bibinfo {year} {2020})}\BibitemShut {NoStop}%
	\bibitem [{\citenamefont {Stadler}\ \emph {et~al.}(2012)\citenamefont
		{Stadler}, \citenamefont {Krinner}, \citenamefont {Meineke}, \citenamefont
		{Brantut},\ and\ \citenamefont {Esslinger}}]{Stadler2012}%
	\BibitemOpen
	\bibfield  {author} {\bibinfo {author} {\bibfnamefont {David}\ \bibnamefont
			{Stadler}}, \bibinfo {author} {\bibfnamefont {Sebastian}\ \bibnamefont
			{Krinner}}, \bibinfo {author} {\bibfnamefont {Jakob}\ \bibnamefont
			{Meineke}}, \bibinfo {author} {\bibfnamefont {Jean-Philippe}\ \bibnamefont
			{Brantut}}, \ and\ \bibinfo {author} {\bibfnamefont {Tilman}\ \bibnamefont
			{Esslinger}},\ }\bibfield  {title} {\enquote {\bibinfo {title} {Observing the
				drop of resistance in the flow of a superfluid fermi gas},}\ }\href {\doibase
		10.1038/nature11613} {\bibfield  {journal} {\bibinfo  {journal} {Nature}\
		}\textbf {\bibinfo {volume} {491}},\ \bibinfo {pages} {736--739} (\bibinfo
		{year} {2012})}\BibitemShut {NoStop}%
	\bibitem [{\citenamefont {Del~Pace}\ \emph {et~al.}(2021)\citenamefont
		{Del~Pace}, \citenamefont {Kwon}, \citenamefont {Zaccanti}, \citenamefont
		{Roati},\ and\ \citenamefont {Scazza}}]{PhysRevLett.126.055301}%
	\BibitemOpen
	\bibfield  {author} {\bibinfo {author} {\bibfnamefont {G.}~\bibnamefont
			{Del~Pace}}, \bibinfo {author} {\bibfnamefont {W.~J.}\ \bibnamefont {Kwon}},
		\bibinfo {author} {\bibfnamefont {M.}~\bibnamefont {Zaccanti}}, \bibinfo
		{author} {\bibfnamefont {G.}~\bibnamefont {Roati}}, \ and\ \bibinfo {author}
		{\bibfnamefont {F.}~\bibnamefont {Scazza}},\ }\bibfield  {title} {\enquote
		{\bibinfo {title} {Tunneling transport of unitary fermions across the
				superfluid transition},}\ }\href {\doibase 10.1103/PhysRevLett.126.055301}
	{\bibfield  {journal} {\bibinfo  {journal} {Phys. Rev. Lett.}\ }\textbf
		{\bibinfo {volume} {126}},\ \bibinfo {pages} {055301} (\bibinfo {year}
		{2021})}\BibitemShut {NoStop}%
	\bibitem [{\citenamefont {Pepino}\ \emph {et~al.}(2010)\citenamefont {Pepino},
		\citenamefont {Cooper}, \citenamefont {Meiser}, \citenamefont {Anderson},\
		and\ \citenamefont {Holland}}]{PRA.82.013640}%
	\BibitemOpen
	\bibfield  {author} {\bibinfo {author} {\bibfnamefont {R.~A.}\ \bibnamefont
			{Pepino}}, \bibinfo {author} {\bibfnamefont {J.}~\bibnamefont {Cooper}},
		\bibinfo {author} {\bibfnamefont {D.}~\bibnamefont {Meiser}}, \bibinfo
		{author} {\bibfnamefont {D.~Z.}\ \bibnamefont {Anderson}}, \ and\ \bibinfo
		{author} {\bibfnamefont {M.~J.}\ \bibnamefont {Holland}},\ }\bibfield
	{title} {\enquote {\bibinfo {title} {Open quantum systems approach to
				atomtronics},}\ }\href {\doibase 10.1103/PhysRevA.82.013640} {\bibfield
		{journal} {\bibinfo  {journal} {Phys. Rev. A}\ }\textbf {\bibinfo {volume}
			{82}},\ \bibinfo {pages} {013640} (\bibinfo {year} {2010})}\BibitemShut
	{NoStop}%
	\bibitem [{\citenamefont {Kapale}\ and\ \citenamefont
		{Dowling}(2005)}]{PhysRevLett.95.173601}%
	\BibitemOpen
	\bibfield  {author} {\bibinfo {author} {\bibfnamefont {Kishore~T.}\
			\bibnamefont {Kapale}}\ and\ \bibinfo {author} {\bibfnamefont {Jonathan~P.}\
			\bibnamefont {Dowling}},\ }\bibfield  {title} {\enquote {\bibinfo {title}
			{Vortex phase qubit: Generating arbitrary, counterrotating, coherent
				superpositions in bose-einstein condensates via optical angular momentum
				beams},}\ }\href {\doibase 10.1103/PhysRevLett.95.173601} {\bibfield
		{journal} {\bibinfo  {journal} {Phys. Rev. Lett.}\ }\textbf {\bibinfo
			{volume} {95}},\ \bibinfo {pages} {173601} (\bibinfo {year}
		{2005})}\BibitemShut {NoStop}%
	\bibitem [{\citenamefont {Hallwood}\ \emph {et~al.}(2006)\citenamefont
		{Hallwood}, \citenamefont {Burnett},\ and\ \citenamefont
		{Dunningham}}]{Hallwood_2006}%
	\BibitemOpen
	\bibfield  {author} {\bibinfo {author} {\bibfnamefont {David~W}\ \bibnamefont
			{Hallwood}}, \bibinfo {author} {\bibfnamefont {Keith}\ \bibnamefont
			{Burnett}}, \ and\ \bibinfo {author} {\bibfnamefont {Jacob}\ \bibnamefont
			{Dunningham}},\ }\bibfield  {title} {\enquote {\bibinfo {title} {Macroscopic
				superpositions of superfluid flows},}\ }\href {\doibase
		10.1088/1367-2630/8/9/180} {\bibfield  {journal} {\bibinfo  {journal} {New
				Journal of Physics}\ }\textbf {\bibinfo {volume} {8}},\ \bibinfo {pages}
		{180} (\bibinfo {year} {2006})}\BibitemShut {NoStop}%
	\bibitem [{\citenamefont {Aghamalyan}\ \emph {et~al.}(2016)\citenamefont
		{Aghamalyan}, \citenamefont {Nguyen}, \citenamefont {Auksztol}, \citenamefont
		{Gan}, \citenamefont {Valado}, \citenamefont {Condylis}, \citenamefont
		{Kwek}, \citenamefont {Dumke},\ and\ \citenamefont
		{Amico}}]{Aghamalyan_2016}%
	\BibitemOpen
	\bibfield  {author} {\bibinfo {author} {\bibfnamefont {D}~\bibnamefont
			{Aghamalyan}}, \bibinfo {author} {\bibfnamefont {N~T}\ \bibnamefont
			{Nguyen}}, \bibinfo {author} {\bibfnamefont {F}~\bibnamefont {Auksztol}},
		\bibinfo {author} {\bibfnamefont {K~S}\ \bibnamefont {Gan}}, \bibinfo
		{author} {\bibfnamefont {M~Martinez}\ \bibnamefont {Valado}}, \bibinfo
		{author} {\bibfnamefont {P~C}\ \bibnamefont {Condylis}}, \bibinfo {author}
		{\bibfnamefont {L-C}\ \bibnamefont {Kwek}}, \bibinfo {author} {\bibfnamefont
			{R}~\bibnamefont {Dumke}}, \ and\ \bibinfo {author} {\bibfnamefont
			{L}~\bibnamefont {Amico}},\ }\bibfield  {title} {\enquote {\bibinfo {title}
			{An atomtronic flux qubit: a ring lattice of bose–einstein condensates
				interrupted by three weak links},}\ }\href {\doibase
		10.1088/1367-2630/18/7/075013} {\bibfield  {journal} {\bibinfo  {journal}
			{New Journal of Physics}\ }\textbf {\bibinfo {volume} {18}},\ \bibinfo
		{pages} {075013} (\bibinfo {year} {2016})}\BibitemShut {NoStop}%
	\bibitem [{\citenamefont {Rubo}\ \emph {et~al.}(2003)\citenamefont {Rubo},
		\citenamefont {Laussy}, \citenamefont {Malpuech}, \citenamefont {Kavokin},\
		and\ \citenamefont {Bigenwald}}]{PhysRevLett.91.156403}%
	\BibitemOpen
	\bibfield  {author} {\bibinfo {author} {\bibfnamefont {Yuri~G.}\ \bibnamefont
			{Rubo}}, \bibinfo {author} {\bibfnamefont {F.~P.}\ \bibnamefont {Laussy}},
		\bibinfo {author} {\bibfnamefont {G.}~\bibnamefont {Malpuech}}, \bibinfo
		{author} {\bibfnamefont {A.}~\bibnamefont {Kavokin}}, \ and\ \bibinfo
		{author} {\bibfnamefont {P.}~\bibnamefont {Bigenwald}},\ }\bibfield  {title}
	{\enquote {\bibinfo {title} {Dynamical theory of polariton amplifiers},}\
	}\href {\doibase 10.1103/PhysRevLett.91.156403} {\bibfield  {journal}
		{\bibinfo  {journal} {Phys. Rev. Lett.}\ }\textbf {\bibinfo {volume} {91}},\
		\bibinfo {pages} {156403} (\bibinfo {year} {2003})}\BibitemShut {NoStop}%
	\bibitem [{\citenamefont {Search}\ and\ \citenamefont
		{Meystre}(2004)}]{PhysRevLett.93.140405}%
	\BibitemOpen
	\bibfield  {author} {\bibinfo {author} {\bibfnamefont {Chris~P.}\
			\bibnamefont {Search}}\ and\ \bibinfo {author} {\bibfnamefont {Pierre}\
			\bibnamefont {Meystre}},\ }\bibfield  {title} {\enquote {\bibinfo {title}
			{Molecular matter-wave amplifier},}\ }\href {\doibase
		10.1103/PhysRevLett.93.140405} {\bibfield  {journal} {\bibinfo  {journal}
			{Phys. Rev. Lett.}\ }\textbf {\bibinfo {volume} {93}},\ \bibinfo {pages}
		{140405} (\bibinfo {year} {2004})}\BibitemShut {NoStop}%
	\bibitem [{\citenamefont {Vaishnav}\ \emph {et~al.}(2008)\citenamefont
		{Vaishnav}, \citenamefont {Ruseckas}, \citenamefont {Clark},\ and\
		\citenamefont {Juzeli\ifmmode~\bar{u}\else
			\={u}\fi{}nas}}]{PhysRevLett.101.265302}%
	\BibitemOpen
	\bibfield  {author} {\bibinfo {author} {\bibfnamefont {J.~Y.}\ \bibnamefont
			{Vaishnav}}, \bibinfo {author} {\bibfnamefont {Julius}\ \bibnamefont
			{Ruseckas}}, \bibinfo {author} {\bibfnamefont {Charles~W.}\ \bibnamefont
			{Clark}}, \ and\ \bibinfo {author} {\bibfnamefont {Gediminas}\ \bibnamefont
			{Juzeli\ifmmode~\bar{u}\else \={u}\fi{}nas}},\ }\bibfield  {title} {\enquote
		{\bibinfo {title} {Spin field effect transistors with ultracold atoms},}\
	}\href {\doibase 10.1103/PhysRevLett.101.265302} {\bibfield  {journal}
		{\bibinfo  {journal} {Phys. Rev. Lett.}\ }\textbf {\bibinfo {volume} {101}},\
		\bibinfo {pages} {265302} (\bibinfo {year} {2008})}\BibitemShut {NoStop}%
	\bibitem [{\citenamefont {Stickney}\ \emph {et~al.}(2007)\citenamefont
		{Stickney}, \citenamefont {Anderson},\ and\ \citenamefont
		{Zozulya}}]{PhysRevA.75.013608}%
	\BibitemOpen
	\bibfield  {author} {\bibinfo {author} {\bibfnamefont {James~A.}\
			\bibnamefont {Stickney}}, \bibinfo {author} {\bibfnamefont {Dana~Z.}\
			\bibnamefont {Anderson}}, \ and\ \bibinfo {author} {\bibfnamefont {Alex~A.}\
			\bibnamefont {Zozulya}},\ }\bibfield  {title} {\enquote {\bibinfo {title}
			{Transistorlike behavior of a bose-einstein condensate in a triple-well
				potential},}\ }\href {\doibase 10.1103/PhysRevA.75.013608} {\bibfield
		{journal} {\bibinfo  {journal} {Phys. Rev. A}\ }\textbf {\bibinfo {volume}
			{75}},\ \bibinfo {pages} {013608} (\bibinfo {year} {2007})}\BibitemShut
	{NoStop}%
	\bibitem [{\citenamefont {Olson}\ \emph {et~al.}(2014)\citenamefont {Olson},
		\citenamefont {Wang}, \citenamefont {Niffenegger}, \citenamefont {Li},
		\citenamefont {Greene},\ and\ \citenamefont {Chen}}]{PhysRevA.90.013616}%
	\BibitemOpen
	\bibfield  {author} {\bibinfo {author} {\bibfnamefont {Abraham~J.}\
			\bibnamefont {Olson}}, \bibinfo {author} {\bibfnamefont {Su-Ju}\ \bibnamefont
			{Wang}}, \bibinfo {author} {\bibfnamefont {Robert~J.}\ \bibnamefont
			{Niffenegger}}, \bibinfo {author} {\bibfnamefont {Chuan-Hsun}\ \bibnamefont
			{Li}}, \bibinfo {author} {\bibfnamefont {Chris~H.}\ \bibnamefont {Greene}}, \
		and\ \bibinfo {author} {\bibfnamefont {Yong~P.}\ \bibnamefont {Chen}},\
	}\bibfield  {title} {\enquote {\bibinfo {title} {Tunable landau-zener
				transitions in a spin-orbit-coupled bose-einstein condensate},}\ }\href
	{\doibase 10.1103/PhysRevA.90.013616} {\bibfield  {journal} {\bibinfo
			{journal} {Phys. Rev. A}\ }\textbf {\bibinfo {volume} {90}},\ \bibinfo
		{pages} {013616} (\bibinfo {year} {2014})}\BibitemShut {NoStop}%
	\bibitem [{\citenamefont {Anderson}(2021)}]{PhysRevA.104.033311}%
	\BibitemOpen
	\bibfield  {author} {\bibinfo {author} {\bibfnamefont {Dana~Z.}\ \bibnamefont
			{Anderson}},\ }\bibfield  {title} {\enquote {\bibinfo {title} {Matter waves,
				single-mode excitations of the matter-wave field, and the atomtronic
				transistor oscillator},}\ }\href {\doibase 10.1103/PhysRevA.104.033311}
	{\bibfield  {journal} {\bibinfo  {journal} {Phys. Rev. A}\ }\textbf {\bibinfo
			{volume} {104}},\ \bibinfo {pages} {033311} (\bibinfo {year}
		{2021})}\BibitemShut {NoStop}%
	\bibitem [{\citenamefont {Caliga}\ \emph {et~al.}(2016)\citenamefont {Caliga},
		\citenamefont {Straatsma}, \citenamefont {Zozulya},\ and\ \citenamefont
		{Anderson}}]{Caliga_2016}%
	\BibitemOpen
	\bibfield  {author} {\bibinfo {author} {\bibfnamefont {Seth~C}\ \bibnamefont
			{Caliga}}, \bibinfo {author} {\bibfnamefont {Cameron J~E}\ \bibnamefont
			{Straatsma}}, \bibinfo {author} {\bibfnamefont {Alex~A}\ \bibnamefont
			{Zozulya}}, \ and\ \bibinfo {author} {\bibfnamefont {Dana~Z}\ \bibnamefont
			{Anderson}},\ }\bibfield  {title} {\enquote {\bibinfo {title} {Principles of
				an atomtronic transistor},}\ }\href {\doibase 10.1088/1367-2630/18/1/015012}
	{\bibfield  {journal} {\bibinfo  {journal} {New Journal of Physics}\ }\textbf
		{\bibinfo {volume} {18}},\ \bibinfo {pages} {015012} (\bibinfo {year}
		{2016})}\BibitemShut {NoStop}%
	\bibitem [{\citenamefont {Shchesnovich}\ and\ \citenamefont
		{Konotop}(2009)}]{PhysRevLett.102.055702}%
	\BibitemOpen
	\bibfield  {author} {\bibinfo {author} {\bibfnamefont {V.~S.}\ \bibnamefont
			{Shchesnovich}}\ and\ \bibinfo {author} {\bibfnamefont {V.~V.}\ \bibnamefont
			{Konotop}},\ }\bibfield  {title} {\enquote {\bibinfo {title} {Quantum
				switching at a mean-field instability of a bose-einstein condensate in an
				optical lattice},}\ }\href {\doibase 10.1103/PhysRevLett.102.055702}
	{\bibfield  {journal} {\bibinfo  {journal} {Phys. Rev. Lett.}\ }\textbf
		{\bibinfo {volume} {102}},\ \bibinfo {pages} {055702} (\bibinfo {year}
		{2009})}\BibitemShut {NoStop}%
	\bibitem [{\citenamefont {Caliga}\ \emph {et~al.}(2017)\citenamefont {Caliga},
		\citenamefont {Straatsma},\ and\ \citenamefont {Anderson}}]{Caliga_2017}%
	\BibitemOpen
	\bibfield  {author} {\bibinfo {author} {\bibfnamefont {Seth~C}\ \bibnamefont
			{Caliga}}, \bibinfo {author} {\bibfnamefont {Cameron J~E}\ \bibnamefont
			{Straatsma}}, \ and\ \bibinfo {author} {\bibfnamefont {Dana~Z}\ \bibnamefont
			{Anderson}},\ }\bibfield  {title} {\enquote {\bibinfo {title} {Experimental
				demonstration of an atomtronic battery},}\ }\href {\doibase
		10.1088/1367-2630/aa56d8} {\bibfield  {journal} {\bibinfo  {journal} {New
				Journal of Physics}\ }\textbf {\bibinfo {volume} {19}},\ \bibinfo {pages}
		{013036} (\bibinfo {year} {2017})}\BibitemShut {NoStop}%
	\bibitem [{\citenamefont {Riedl}\ \emph {et~al.}(2012)\citenamefont {Riedl},
		\citenamefont {Lettner}, \citenamefont {Vo}, \citenamefont {Baur},
		\citenamefont {Rempe},\ and\ \citenamefont {D\"urr}}]{PhysRevA.85.022318}%
	\BibitemOpen
	\bibfield  {author} {\bibinfo {author} {\bibfnamefont {Stefan}\ \bibnamefont
			{Riedl}}, \bibinfo {author} {\bibfnamefont {Matthias}\ \bibnamefont
			{Lettner}}, \bibinfo {author} {\bibfnamefont {Christoph}\ \bibnamefont {Vo}},
		\bibinfo {author} {\bibfnamefont {Simon}\ \bibnamefont {Baur}}, \bibinfo
		{author} {\bibfnamefont {Gerhard}\ \bibnamefont {Rempe}}, \ and\ \bibinfo
		{author} {\bibfnamefont {Stephan}\ \bibnamefont {D\"urr}},\ }\bibfield
	{title} {\enquote {\bibinfo {title} {Bose-einstein condensate as a quantum
				memory for a photonic polarization qubit},}\ }\href {\doibase
		10.1103/PhysRevA.85.022318} {\bibfield  {journal} {\bibinfo  {journal} {Phys.
				Rev. A}\ }\textbf {\bibinfo {volume} {85}},\ \bibinfo {pages} {022318}
		(\bibinfo {year} {2012})}\BibitemShut {NoStop}%
	\bibitem [{\citenamefont {Da~Ros}\ \emph {et~al.}(2023)\citenamefont {Da~Ros},
		\citenamefont {Kanthak}, \citenamefont {Sa\ifmmode~\breve{g}\else
			\u{g}\fi{}lamy\"urek}, \citenamefont {G\"undo\ifmmode~\breve{g}\else
			\u{g}\fi{}an},\ and\ \citenamefont {Krutzik}}]{PhysRevResearch.5.033003}%
	\BibitemOpen
	\bibfield  {author} {\bibinfo {author} {\bibfnamefont {Elisa}\ \bibnamefont
			{Da~Ros}}, \bibinfo {author} {\bibfnamefont {Simon}\ \bibnamefont {Kanthak}},
		\bibinfo {author} {\bibfnamefont {Erhan}\ \bibnamefont
			{Sa\ifmmode~\breve{g}\else \u{g}\fi{}lamy\"urek}}, \bibinfo {author}
		{\bibfnamefont {Mustafa}\ \bibnamefont {G\"undo\ifmmode~\breve{g}\else
				\u{g}\fi{}an}}, \ and\ \bibinfo {author} {\bibfnamefont {Markus}\
			\bibnamefont {Krutzik}},\ }\bibfield  {title} {\enquote {\bibinfo {title}
			{Proposal for a long-lived quantum memory using matter-wave optics with
				bose-einstein condensates in microgravity},}\ }\href {\doibase
		10.1103/PhysRevResearch.5.033003} {\bibfield  {journal} {\bibinfo  {journal}
			{Phys. Rev. Res.}\ }\textbf {\bibinfo {volume} {5}},\ \bibinfo {pages}
		{033003} (\bibinfo {year} {2023})}\BibitemShut {NoStop}%
	\bibitem [{\citenamefont {Saglamyurek}\ \emph {et~al.}(2021)\citenamefont
		{Saglamyurek}, \citenamefont {Hrushevskyi}, \citenamefont {Rastogi},
		\citenamefont {Cooke}, \citenamefont {Smith},\ and\ \citenamefont
		{LeBlanc}}]{NJP.23.043028}%
	\BibitemOpen
	\bibfield  {author} {\bibinfo {author} {\bibfnamefont {Erhan}\ \bibnamefont
			{Saglamyurek}}, \bibinfo {author} {\bibfnamefont {Taras}\ \bibnamefont
			{Hrushevskyi}}, \bibinfo {author} {\bibfnamefont {Anindya}\ \bibnamefont
			{Rastogi}}, \bibinfo {author} {\bibfnamefont {Logan~W}\ \bibnamefont
			{Cooke}}, \bibinfo {author} {\bibfnamefont {Benjamin~D}\ \bibnamefont
			{Smith}}, \ and\ \bibinfo {author} {\bibfnamefont {Lindsay~J}\ \bibnamefont
			{LeBlanc}},\ }\bibfield  {title} {\enquote {\bibinfo {title} {Storing short
				single-photon-level optical pulses in bose–einstein condensates for
				high-performance quantum memory},}\ }\href {\doibase
		10.1088/1367-2630/abf1d9} {\bibfield  {journal} {\bibinfo  {journal} {New
				Journal of Physics}\ }\textbf {\bibinfo {volume} {23}},\ \bibinfo {pages}
		{043028} (\bibinfo {year} {2021})}\BibitemShut {NoStop}%
	\bibitem [{\citenamefont {Pepino}\ \emph {et~al.}(2009)\citenamefont {Pepino},
		\citenamefont {Cooper}, \citenamefont {Anderson},\ and\ \citenamefont
		{Holland}}]{PhysRevLett.103.140405}%
	\BibitemOpen
	\bibfield  {author} {\bibinfo {author} {\bibfnamefont {R.~A.}\ \bibnamefont
			{Pepino}}, \bibinfo {author} {\bibfnamefont {J.}~\bibnamefont {Cooper}},
		\bibinfo {author} {\bibfnamefont {D.~Z.}\ \bibnamefont {Anderson}}, \ and\
		\bibinfo {author} {\bibfnamefont {M.~J.}\ \bibnamefont {Holland}},\
	}\bibfield  {title} {\enquote {\bibinfo {title} {Atomtronic circuits of
				diodes and transistors},}\ }\href {\doibase 10.1103/PhysRevLett.103.140405}
	{\bibfield  {journal} {\bibinfo  {journal} {Phys. Rev. Lett.}\ }\textbf
		{\bibinfo {volume} {103}},\ \bibinfo {pages} {140405} (\bibinfo {year}
		{2009})}\BibitemShut {NoStop}%
	\bibitem [{\citenamefont {Seaman}\ \emph {et~al.}(2007)\citenamefont {Seaman},
		\citenamefont {Kr\"amer}, \citenamefont {Anderson},\ and\ \citenamefont
		{Holland}}]{PhysRevA.75.023615}%
	\BibitemOpen
	\bibfield  {author} {\bibinfo {author} {\bibfnamefont {B.~T.}\ \bibnamefont
			{Seaman}}, \bibinfo {author} {\bibfnamefont {M.}~\bibnamefont {Kr\"amer}},
		\bibinfo {author} {\bibfnamefont {D.~Z.}\ \bibnamefont {Anderson}}, \ and\
		\bibinfo {author} {\bibfnamefont {M.~J.}\ \bibnamefont {Holland}},\
	}\bibfield  {title} {\enquote {\bibinfo {title} {Atomtronics: Ultracold-atom
				analogs of electronic devices},}\ }\href {\doibase
		10.1103/PhysRevA.75.023615} {\bibfield  {journal} {\bibinfo  {journal} {Phys.
				Rev. A}\ }\textbf {\bibinfo {volume} {75}},\ \bibinfo {pages} {023615}
		(\bibinfo {year} {2007})}\BibitemShut {NoStop}%
	\bibitem [{\citenamefont {Ruschhaupt}\ and\ \citenamefont
		{Muga}(2004)}]{PhysRevA.70.061604}%
	\BibitemOpen
	\bibfield  {author} {\bibinfo {author} {\bibfnamefont {A.}~\bibnamefont
			{Ruschhaupt}}\ and\ \bibinfo {author} {\bibfnamefont {J.~G.}\ \bibnamefont
			{Muga}},\ }\bibfield  {title} {\enquote {\bibinfo {title} {Atom diode: A
				laser device for a unidirectional transmission of ground-state atoms},}\
	}\href {\doibase 10.1103/PhysRevA.70.061604} {\bibfield  {journal} {\bibinfo
			{journal} {Phys. Rev. A}\ }\textbf {\bibinfo {volume} {70}},\ \bibinfo
		{pages} {061604} (\bibinfo {year} {2004})}\BibitemShut {NoStop}%
	\bibitem [{\citenamefont {Madasu}\ \emph {et~al.}(2022)\citenamefont {Madasu},
		\citenamefont {Hasan}, \citenamefont {Rathod}, \citenamefont {Kwong},\ and\
		\citenamefont {Wilkowski}}]{PhysRevResearch.4.033180}%
	\BibitemOpen
	\bibfield  {author} {\bibinfo {author} {\bibfnamefont {Chetan~Sriram}\
			\bibnamefont {Madasu}}, \bibinfo {author} {\bibfnamefont {Mehedi}\
			\bibnamefont {Hasan}}, \bibinfo {author} {\bibfnamefont {Ketan~Damji}\
			\bibnamefont {Rathod}}, \bibinfo {author} {\bibfnamefont {Chang~Chi}\
			\bibnamefont {Kwong}}, \ and\ \bibinfo {author} {\bibfnamefont {David}\
			\bibnamefont {Wilkowski}},\ }\bibfield  {title} {\enquote {\bibinfo {title}
			{Datta-das transistor for atomtronic circuits using artificial gauge
				fields},}\ }\href {\doibase 10.1103/PhysRevResearch.4.033180} {\bibfield
		{journal} {\bibinfo  {journal} {Phys. Rev. Res.}\ }\textbf {\bibinfo {volume}
			{4}},\ \bibinfo {pages} {033180} (\bibinfo {year} {2022})}\BibitemShut
	{NoStop}%
	\bibitem [{\citenamefont {Ryu}\ \emph {et~al.}(2013)\citenamefont {Ryu},
		\citenamefont {Blackburn}, \citenamefont {Blinova},\ and\ \citenamefont
		{Boshier}}]{PhysRevLett.111.205301}%
	\BibitemOpen
	\bibfield  {author} {\bibinfo {author} {\bibfnamefont {C.}~\bibnamefont
			{Ryu}}, \bibinfo {author} {\bibfnamefont {P.~W.}\ \bibnamefont {Blackburn}},
		\bibinfo {author} {\bibfnamefont {A.~A.}\ \bibnamefont {Blinova}}, \ and\
		\bibinfo {author} {\bibfnamefont {M.~G.}\ \bibnamefont {Boshier}},\
	}\bibfield  {title} {\enquote {\bibinfo {title} {Experimental realization of
				josephson junctions for an atom squid},}\ }\href {\doibase
		10.1103/PhysRevLett.111.205301} {\bibfield  {journal} {\bibinfo  {journal}
			{Phys. Rev. Lett.}\ }\textbf {\bibinfo {volume} {111}},\ \bibinfo {pages}
		{205301} (\bibinfo {year} {2013})}\BibitemShut {NoStop}%
	\bibitem [{\citenamefont {Lau}\ \emph {et~al.}(2023)\citenamefont {Lau},
		\citenamefont {Gan}, \citenamefont {Dumke}, \citenamefont {Amico},
		\citenamefont {Kwek},\ and\ \citenamefont {Haug}}]{PhysRevA.107.L051303}%
	\BibitemOpen
	\bibfield  {author} {\bibinfo {author} {\bibfnamefont {Jonathan Wei~Zhong}\
			\bibnamefont {Lau}}, \bibinfo {author} {\bibfnamefont {Koon~Siang}\
			\bibnamefont {Gan}}, \bibinfo {author} {\bibfnamefont {Rainer}\ \bibnamefont
			{Dumke}}, \bibinfo {author} {\bibfnamefont {Luigi}\ \bibnamefont {Amico}},
		\bibinfo {author} {\bibfnamefont {Leong-Chuan}\ \bibnamefont {Kwek}}, \ and\
		\bibinfo {author} {\bibfnamefont {Tobias}\ \bibnamefont {Haug}},\ }\bibfield
	{title} {\enquote {\bibinfo {title} {Atomtronic multiterminal aharonov-bohm
				interferometer},}\ }\href {\doibase 10.1103/PhysRevA.107.L051303} {\bibfield
		{journal} {\bibinfo  {journal} {Phys. Rev. A}\ }\textbf {\bibinfo {volume}
			{107}},\ \bibinfo {pages} {L051303} (\bibinfo {year} {2023})}\BibitemShut
	{NoStop}%
	\bibitem [{\citenamefont {Pezz\`e}\ \emph {et~al.}(2018)\citenamefont
		{Pezz\`e}, \citenamefont {Smerzi}, \citenamefont {Oberthaler}, \citenamefont
		{Schmied},\ and\ \citenamefont {Treutlein}}]{RevModPhys.90.035005}%
	\BibitemOpen
	\bibfield  {author} {\bibinfo {author} {\bibfnamefont {Luca}\ \bibnamefont
			{Pezz\`e}}, \bibinfo {author} {\bibfnamefont {Augusto}\ \bibnamefont
			{Smerzi}}, \bibinfo {author} {\bibfnamefont {Markus~K.}\ \bibnamefont
			{Oberthaler}}, \bibinfo {author} {\bibfnamefont {Roman}\ \bibnamefont
			{Schmied}}, \ and\ \bibinfo {author} {\bibfnamefont {Philipp}\ \bibnamefont
			{Treutlein}},\ }\bibfield  {title} {\enquote {\bibinfo {title} {Quantum
				metrology with nonclassical states of atomic ensembles},}\ }\href {\doibase
		10.1103/RevModPhys.90.035005} {\bibfield  {journal} {\bibinfo  {journal}
			{Rev. Mod. Phys.}\ }\textbf {\bibinfo {volume} {90}},\ \bibinfo {pages}
		{035005} (\bibinfo {year} {2018})}\BibitemShut {NoStop}%
	\bibitem [{\citenamefont {Haroche}\ and\ \citenamefont
		{Raimond}(2006)}]{Haroche}%
	\BibitemOpen
	\bibfield  {author} {\bibinfo {author} {\bibfnamefont {S.}~\bibnamefont
			{Haroche}}\ and\ \bibinfo {author} {\bibfnamefont {J.-M.}\ \bibnamefont
			{Raimond}},\ }\href@noop {} {\emph {\bibinfo {title} {Exploring the Quantum:
				Atoms, Cavities, and Photons}}}\ (\bibinfo  {publisher} {Oxford University
		Press, New York},\ \bibinfo {year} {2006})\BibitemShut {NoStop}%
	\bibitem [{\citenamefont {Luick}\ \emph {et~al.}(2020)\citenamefont {Luick},
		\citenamefont {Sobirey}, \citenamefont {Bohlen}, \citenamefont {Singh},
		\citenamefont {Mathey}, \citenamefont {Lompe},\ and\ \citenamefont
		{Moritz}}]{doi:10.1126/science.aaz2342}%
	\BibitemOpen
	\bibfield  {author} {\bibinfo {author} {\bibfnamefont {Niclas}\ \bibnamefont
			{Luick}}, \bibinfo {author} {\bibfnamefont {Lennart}\ \bibnamefont
			{Sobirey}}, \bibinfo {author} {\bibfnamefont {Markus}\ \bibnamefont
			{Bohlen}}, \bibinfo {author} {\bibfnamefont {Vijay~Pal}\ \bibnamefont
			{Singh}}, \bibinfo {author} {\bibfnamefont {Ludwig}\ \bibnamefont {Mathey}},
		\bibinfo {author} {\bibfnamefont {Thomas}\ \bibnamefont {Lompe}}, \ and\
		\bibinfo {author} {\bibfnamefont {Henning}\ \bibnamefont {Moritz}},\
	}\bibfield  {title} {\enquote {\bibinfo {title} {An ideal josephson junction
				in an ultracold two-dimensional fermi gas},}\ }\href {\doibase
		10.1126/science.aaz2342} {\bibfield  {journal} {\bibinfo  {journal}
			{Science}\ }\textbf {\bibinfo {volume} {369}},\ \bibinfo {pages} {89--91}
		(\bibinfo {year} {2020})}\BibitemShut {NoStop}%
	\bibitem [{\citenamefont {Papoular}\ \emph {et~al.}(2016)\citenamefont
		{Papoular}, \citenamefont {Pitaevskii},\ and\ \citenamefont
		{Stringari}}]{PhysRevA.94.023622}%
	\BibitemOpen
	\bibfield  {author} {\bibinfo {author} {\bibfnamefont {D.~J.}\ \bibnamefont
			{Papoular}}, \bibinfo {author} {\bibfnamefont {L.~P.}\ \bibnamefont
			{Pitaevskii}}, \ and\ \bibinfo {author} {\bibfnamefont {S.}~\bibnamefont
			{Stringari}},\ }\bibfield  {title} {\enquote {\bibinfo {title} {Quantized
				conductance through the quantum evaporation of bosonic atoms},}\ }\href
	{\doibase 10.1103/PhysRevA.94.023622} {\bibfield  {journal} {\bibinfo
			{journal} {Phys. Rev. A}\ }\textbf {\bibinfo {volume} {94}},\ \bibinfo
		{pages} {023622} (\bibinfo {year} {2016})}\BibitemShut {NoStop}%
	\bibitem [{\citenamefont {Liu}\ \emph {et~al.}(2017)\citenamefont {Liu},
		\citenamefont {Zhai},\ and\ \citenamefont {Zhang}}]{PhysRevA.95.013623}%
	\BibitemOpen
	\bibfield  {author} {\bibinfo {author} {\bibfnamefont {Boyang}\ \bibnamefont
			{Liu}}, \bibinfo {author} {\bibfnamefont {Hui}\ \bibnamefont {Zhai}}, \ and\
		\bibinfo {author} {\bibfnamefont {Shizhong}\ \bibnamefont {Zhang}},\
	}\bibfield  {title} {\enquote {\bibinfo {title} {Anomalous conductance of a
				strongly interacting fermi gas through a quantum point contact},}\ }\href
	{\doibase 10.1103/PhysRevA.95.013623} {\bibfield  {journal} {\bibinfo
			{journal} {Phys. Rev. A}\ }\textbf {\bibinfo {volume} {95}},\ \bibinfo
		{pages} {013623} (\bibinfo {year} {2017})}\BibitemShut {NoStop}%
	\bibitem [{\citenamefont {Brantut}\ \emph {et~al.}(2013)\citenamefont
		{Brantut}, \citenamefont {Grenier}, \citenamefont {Meineke}, \citenamefont
		{Stadler}, \citenamefont {Krinner}, \citenamefont {Kollath}, \citenamefont
		{Esslinger},\ and\ \citenamefont {Georges}}]{doi:10.1126/science.1242308}%
	\BibitemOpen
	\bibfield  {author} {\bibinfo {author} {\bibfnamefont {Jean-Philippe}\
			\bibnamefont {Brantut}}, \bibinfo {author} {\bibfnamefont {Charles}\
			\bibnamefont {Grenier}}, \bibinfo {author} {\bibfnamefont {Jakob}\
			\bibnamefont {Meineke}}, \bibinfo {author} {\bibfnamefont {David}\
			\bibnamefont {Stadler}}, \bibinfo {author} {\bibfnamefont {Sebastian}\
			\bibnamefont {Krinner}}, \bibinfo {author} {\bibfnamefont {Corinna}\
			\bibnamefont {Kollath}}, \bibinfo {author} {\bibfnamefont {Tilman}\
			\bibnamefont {Esslinger}}, \ and\ \bibinfo {author} {\bibfnamefont {Antoine}\
			\bibnamefont {Georges}},\ }\bibfield  {title} {\enquote {\bibinfo {title} {A
				thermoelectric heat engine with ultracold atoms},}\ }\href {\doibase
		10.1126/science.1242308} {\bibfield  {journal} {\bibinfo  {journal}
			{Science}\ }\textbf {\bibinfo {volume} {342}},\ \bibinfo {pages} {713--715}
		(\bibinfo {year} {2013})} \BibitemShut	{NoStop}%
	\bibitem [{\citenamefont {Pigneur}\ \emph {et~al.}(2018)\citenamefont
		{Pigneur}, \citenamefont {Berrada}, \citenamefont {Bonneau}, \citenamefont
		{Schumm}, \citenamefont {Demler},\ and\ \citenamefont
		{Schmiedmayer}}]{PhysRevLett.120.173601}%
	\BibitemOpen
	\bibfield  {author} {\bibinfo {author} {\bibfnamefont {Marine}\ \bibnamefont
			{Pigneur}}, \bibinfo {author} {\bibfnamefont {Tarik}\ \bibnamefont
			{Berrada}}, \bibinfo {author} {\bibfnamefont {Marie}\ \bibnamefont
			{Bonneau}}, \bibinfo {author} {\bibfnamefont {Thorsten}\ \bibnamefont
			{Schumm}}, \bibinfo {author} {\bibfnamefont {Eugene}\ \bibnamefont {Demler}},
		\ and\ \bibinfo {author} {\bibfnamefont {J\"org}\ \bibnamefont
			{Schmiedmayer}},\ }\bibfield  {title} {\enquote {\bibinfo {title} {Relaxation
				to a phase-locked equilibrium state in a one-dimensional bosonic josephson
				junction},}\ }\href {\doibase 10.1103/PhysRevLett.120.173601} {\bibfield
		{journal} {\bibinfo  {journal} {Phys. Rev. Lett.}\ }\textbf {\bibinfo
			{volume} {120}},\ \bibinfo {pages} {173601} (\bibinfo {year}
		{2018})}\BibitemShut {NoStop}%
	\bibitem [{\citenamefont {Gauthier}\ \emph {et~al.}(2019)\citenamefont
		{Gauthier}, \citenamefont {Szigeti}, \citenamefont {Reeves}, \citenamefont
		{Baker}, \citenamefont {Bell}, \citenamefont {Rubinsztein-Dunlop},
		\citenamefont {Davis},\ and\ \citenamefont {Neely}}]{PhysRevLett.123.260402}%
	\BibitemOpen
	\bibfield  {author} {\bibinfo {author} {\bibfnamefont {Guillaume}\
			\bibnamefont {Gauthier}}, \bibinfo {author} {\bibfnamefont {Stuart~S.}\
			\bibnamefont {Szigeti}}, \bibinfo {author} {\bibfnamefont {Matthew~T.}\
			\bibnamefont {Reeves}}, \bibinfo {author} {\bibfnamefont {Mark}\ \bibnamefont
			{Baker}}, \bibinfo {author} {\bibfnamefont {Thomas~A.}\ \bibnamefont {Bell}},
		\bibinfo {author} {\bibfnamefont {Halina}\ \bibnamefont
			{Rubinsztein-Dunlop}}, \bibinfo {author} {\bibfnamefont {Matthew~J.}\
			\bibnamefont {Davis}}, \ and\ \bibinfo {author} {\bibfnamefont {Tyler~W.}\
			\bibnamefont {Neely}},\ }\bibfield  {title} {\enquote {\bibinfo {title}
			{Quantitative acoustic models for superfluid circuits},}\ }\href {\doibase
		10.1103/PhysRevLett.123.260402} {\bibfield  {journal} {\bibinfo  {journal}
			{Phys. Rev. Lett.}\ }\textbf {\bibinfo {volume} {123}},\ \bibinfo {pages}
		{260402} (\bibinfo {year} {2019})}\BibitemShut {NoStop}%
	\bibitem [{\citenamefont {Eckel}\ \emph {et~al.}(2016)\citenamefont {Eckel},
		\citenamefont {Lee}, \citenamefont {Jendrzejewski}, \citenamefont {Lobb},
		\citenamefont {Campbell},\ and\ \citenamefont {Hill}}]{PhysRevA.93.063619}%
	\BibitemOpen
	\bibfield  {author} {\bibinfo {author} {\bibfnamefont {S.}~\bibnamefont
			{Eckel}}, \bibinfo {author} {\bibfnamefont {Jeffrey~G.}\ \bibnamefont {Lee}},
		\bibinfo {author} {\bibfnamefont {F.}~\bibnamefont {Jendrzejewski}}, \bibinfo
		{author} {\bibfnamefont {C.~J.}\ \bibnamefont {Lobb}}, \bibinfo {author}
		{\bibfnamefont {G.~K.}\ \bibnamefont {Campbell}}, \ and\ \bibinfo {author}
		{\bibfnamefont {W.~T.}\ \bibnamefont {Hill}},\ }\bibfield  {title} {\enquote
		{\bibinfo {title} {Contact resistance and phase slips in mesoscopic
				superfluid-atom transport},}\ }\href {\doibase 10.1103/PhysRevA.93.063619}
	{\bibfield  {journal} {\bibinfo  {journal} {Phys. Rev. A}\ }\textbf {\bibinfo
			{volume} {93}},\ \bibinfo {pages} {063619} (\bibinfo {year}
		{2016})}\BibitemShut {NoStop}%
	\bibitem [{\citenamefont {Shunyaev}\ \emph {et~al.}(2016)\citenamefont
		{Shunyaev}, \citenamefont {Elistratov},\ and\ \citenamefont
		{Lozovik}}]{PhysRevA.94.053625}%
	\BibitemOpen
	\bibfield  {author} {\bibinfo {author} {\bibfnamefont {I.~V.}\ \bibnamefont
			{Shunyaev}}, \bibinfo {author} {\bibfnamefont {A.~A.}\ \bibnamefont
			{Elistratov}}, \ and\ \bibinfo {author} {\bibfnamefont {Yu.~E.}\ \bibnamefont
			{Lozovik}},\ }\bibfield  {title} {\enquote {\bibinfo {title} {Bose-einstein
				condensates and the spectrum of excitations in a two-dimensional channel},}\
	}\href {\doibase 10.1103/PhysRevA.94.053625} {\bibfield  {journal} {\bibinfo
			{journal} {Phys. Rev. A}\ }\textbf {\bibinfo {volume} {94}},\ \bibinfo
		{pages} {053625} (\bibinfo {year} {2016})}\BibitemShut {NoStop}%
	\bibitem [{\citenamefont {Robinson}(1951)}]{PhysRev.82.440}%
	\BibitemOpen
	\bibfield  {author} {\bibinfo {author} {\bibfnamefont {John~E.}\ \bibnamefont
			{Robinson}},\ }\bibfield  {title} {\enquote {\bibinfo {title} {Adiabatic
				oscillations in liquid helium},}\ }\href {\doibase 10.1103/PhysRev.82.440}
	{\bibfield  {journal} {\bibinfo  {journal} {Phys. Rev.}\ }\textbf {\bibinfo
			{volume} {82}},\ \bibinfo {pages} {440--441} (\bibinfo {year}
		{1951})}\BibitemShut {NoStop}%
	\bibitem [{\citenamefont {Li}\ \emph {et~al.}(2016)\citenamefont {Li},
		\citenamefont {Eckel}, \citenamefont {Eller}, \citenamefont {Warren},
		\citenamefont {Clark},\ and\ \citenamefont {Edwards}}]{PhysRevA.94.023626}%
	\BibitemOpen
	\bibfield  {author} {\bibinfo {author} {\bibfnamefont {Aijun}\ \bibnamefont
			{Li}}, \bibinfo {author} {\bibfnamefont {Stephen}\ \bibnamefont {Eckel}},
		\bibinfo {author} {\bibfnamefont {Benjamin}\ \bibnamefont {Eller}}, \bibinfo
		{author} {\bibfnamefont {Kayla~E.}\ \bibnamefont {Warren}}, \bibinfo {author}
		{\bibfnamefont {Charles~W.}\ \bibnamefont {Clark}}, \ and\ \bibinfo {author}
		{\bibfnamefont {Mark}\ \bibnamefont {Edwards}},\ }\bibfield  {title}
	{\enquote {\bibinfo {title} {Superfluid transport dynamics in a capacitive
				atomtronic circuit},}\ }\href {\doibase 10.1103/PhysRevA.94.023626}
	{\bibfield  {journal} {\bibinfo  {journal} {Phys. Rev. A}\ }\textbf {\bibinfo
			{volume} {94}},\ \bibinfo {pages} {023626} (\bibinfo {year}
		{2016})}\BibitemShut {NoStop}%
	\bibitem [{\citenamefont {Burchianti}\ \emph {et~al.}(2018)\citenamefont
		{Burchianti}, \citenamefont {Scazza}, \citenamefont {Amico}, \citenamefont
		{Valtolina}, \citenamefont {Seman}, \citenamefont {Fort}, \citenamefont
		{Zaccanti}, \citenamefont {Inguscio},\ and\ \citenamefont
		{Roati}}]{PhysRevLett.120.025302}%
	\BibitemOpen
	\bibfield  {author} {\bibinfo {author} {\bibfnamefont {A.}~\bibnamefont
			{Burchianti}}, \bibinfo {author} {\bibfnamefont {F.}~\bibnamefont {Scazza}},
		\bibinfo {author} {\bibfnamefont {A.}~\bibnamefont {Amico}}, \bibinfo
		{author} {\bibfnamefont {G.}~\bibnamefont {Valtolina}}, \bibinfo {author}
		{\bibfnamefont {J.~A.}\ \bibnamefont {Seman}}, \bibinfo {author}
		{\bibfnamefont {C.}~\bibnamefont {Fort}}, \bibinfo {author} {\bibfnamefont
			{M.}~\bibnamefont {Zaccanti}}, \bibinfo {author} {\bibfnamefont
			{M.}~\bibnamefont {Inguscio}}, \ and\ \bibinfo {author} {\bibfnamefont
			{G.}~\bibnamefont {Roati}},\ }\bibfield  {title} {\enquote {\bibinfo {title}
			{Connecting dissipation and phase slips in a josephson junction between
				fermionic superfluids},}\ }\href {\doibase 10.1103/PhysRevLett.120.025302}
	{\bibfield  {journal} {\bibinfo  {journal} {Phys. Rev. Lett.}\ }\textbf
		{\bibinfo {volume} {120}},\ \bibinfo {pages} {025302} (\bibinfo {year}
		{2018})}\BibitemShut {NoStop}%
	\bibitem [{\citenamefont {Xhani}\ \emph {et~al.}(2020)\citenamefont {Xhani},
		\citenamefont {Neri}, \citenamefont {Galantucci}, \citenamefont {Scazza},
		\citenamefont {Burchianti}, \citenamefont {Lee}, \citenamefont {Barenghi},
		\citenamefont {Trombettoni}, \citenamefont {Inguscio}, \citenamefont
		{Zaccanti}, \citenamefont {Roati},\ and\ \citenamefont
		{Proukakis}}]{PhysRevLett.124.045301}%
	\BibitemOpen
	\bibfield  {author} {\bibinfo {author} {\bibfnamefont {K.}~\bibnamefont
			{Xhani}}, \bibinfo {author} {\bibfnamefont {E.}~\bibnamefont {Neri}},
		\bibinfo {author} {\bibfnamefont {L.}~\bibnamefont {Galantucci}}, \bibinfo
		{author} {\bibfnamefont {F.}~\bibnamefont {Scazza}}, \bibinfo {author}
		{\bibfnamefont {A.}~\bibnamefont {Burchianti}}, \bibinfo {author}
		{\bibfnamefont {K.-L.}\ \bibnamefont {Lee}}, \bibinfo {author} {\bibfnamefont
			{C.~F.}\ \bibnamefont {Barenghi}}, \bibinfo {author} {\bibfnamefont
			{A.}~\bibnamefont {Trombettoni}}, \bibinfo {author} {\bibfnamefont
			{M.}~\bibnamefont {Inguscio}}, \bibinfo {author} {\bibfnamefont
			{M.}~\bibnamefont {Zaccanti}}, \bibinfo {author} {\bibfnamefont
			{G.}~\bibnamefont {Roati}}, \ and\ \bibinfo {author} {\bibfnamefont {N.~P.}\
			\bibnamefont {Proukakis}},\ }\bibfield  {title} {\enquote {\bibinfo {title}
			{Critical transport and vortex dynamics in a thin atomic josephson
				junction},}\ }\href {\doibase 10.1103/PhysRevLett.124.045301} {\bibfield
		{journal} {\bibinfo  {journal} {Phys. Rev. Lett.}\ }\textbf {\bibinfo
			{volume} {124}},\ \bibinfo {pages} {045301} (\bibinfo {year}
		{2020})}\BibitemShut {NoStop}%
	\bibitem [{\citenamefont {Valtolina}\ \emph {et~al.}(2015)\citenamefont
		{Valtolina}, \citenamefont {Burchianti}, \citenamefont {Amico}, \citenamefont
		{Neri}, \citenamefont {Xhani}, \citenamefont {Seman}, \citenamefont
		{Trombettoni}, \citenamefont {Smerzi}, \citenamefont {Zaccanti},
		\citenamefont {Inguscio},\ and\ \citenamefont {Roati}}]{Science.aac9725}%
	\BibitemOpen
	\bibfield  {author} {\bibinfo {author} {\bibfnamefont {Giacomo}\ \bibnamefont
			{Valtolina}}, \bibinfo {author} {\bibfnamefont {Alessia}\ \bibnamefont
			{Burchianti}}, \bibinfo {author} {\bibfnamefont {Andrea}\ \bibnamefont
			{Amico}}, \bibinfo {author} {\bibfnamefont {Elettra}\ \bibnamefont {Neri}},
		\bibinfo {author} {\bibfnamefont {Klejdja}\ \bibnamefont {Xhani}}, \bibinfo
		{author} {\bibfnamefont {Jorge~Amin}\ \bibnamefont {Seman}}, \bibinfo
		{author} {\bibfnamefont {Andrea}\ \bibnamefont {Trombettoni}}, \bibinfo
		{author} {\bibfnamefont {Augusto}\ \bibnamefont {Smerzi}}, \bibinfo {author}
		{\bibfnamefont {Matteo}\ \bibnamefont {Zaccanti}}, \bibinfo {author}
		{\bibfnamefont {Massimo}\ \bibnamefont {Inguscio}}, \ and\ \bibinfo {author}
		{\bibfnamefont {Giacomo}\ \bibnamefont {Roati}},\ }\bibfield  {title}
	{\enquote {\bibinfo {title} {Josephson effect in fermionic superfluids across
				the bec-bcs crossover},}\ }\href {\doibase 10.1126/science.aac9725}
	{\bibfield  {journal} {\bibinfo  {journal} {Science}\ }\textbf {\bibinfo
			{volume} {350}},\ \bibinfo {pages} {1505--1508} (\bibinfo {year} {2015})}\BibitemShut {NoStop}%
	\bibitem [{\citenamefont {Yang}\ \emph {et~al.}(2013)\citenamefont {Yang},
		\citenamefont {Xiong},\ and\ \citenamefont {Benedict}}]{PRA.87.023603}%
	\BibitemOpen
	\bibfield  {author} {\bibinfo {author} {\bibfnamefont {T.}~\bibnamefont
			{Yang}}, \bibinfo {author} {\bibfnamefont {B.}~\bibnamefont {Xiong}}, \ and\
		\bibinfo {author} {\bibfnamefont {Keith~A.}\ \bibnamefont {Benedict}},\
	}\bibfield  {title} {\enquote {\bibinfo {title} {Dynamical excitations in the
				collision of two-dimensional bose-einstein condensates},}\ }\href {\doibase
		10.1103/PhysRevA.87.023603} {\bibfield  {journal} {\bibinfo  {journal} {Phys.
				Rev. A}\ }\textbf {\bibinfo {volume} {87}},\ \bibinfo {pages} {023603}
		(\bibinfo {year} {2013})}\BibitemShut {NoStop}%
	\bibitem [{\citenamefont {Yang}\ \emph {et~al.}(2016)\citenamefont {Yang},
		\citenamefont {Hu}, \citenamefont {Zou},\ and\ \citenamefont
		{Liu}}]{SR.6.29066}%
	\BibitemOpen
	\bibfield  {author} {\bibinfo {author} {\bibfnamefont {Tao}\ \bibnamefont
			{Yang}}, \bibinfo {author} {\bibfnamefont {Zhi-Qiang}\ \bibnamefont {Hu}},
		\bibinfo {author} {\bibfnamefont {Shan}\ \bibnamefont {Zou}}, \ and\ \bibinfo
		{author} {\bibfnamefont {Wu-Ming}\ \bibnamefont {Liu}},\ }\bibfield  {title}
	{\enquote {\bibinfo {title} {Dynamics of vortex quadrupoles in nonrotating
				trapped {Bose-Einstein} condensates},}\ }\href
	{https://linkspringer.53yu.com/content/pdf/10.1038/srep29066.pdf} {\bibfield
		{journal} {\bibinfo  {journal} {Scientific Reports}\ }\textbf {\bibinfo
			{volume} {6}},\ \bibinfo {pages} {29066} (\bibinfo {year}
		{2016})}\BibitemShut {NoStop}%
	\bibitem [{\citenamefont {Xing}\ \emph {et~al.}(2023)\citenamefont {Xing},
		\citenamefont {Bai}, \citenamefont {Xiong}, \citenamefont {Zheng},\ and\
		\citenamefont {Yang}}]{FOP.18.62302}%
	\BibitemOpen
	\bibfield  {author} {\bibinfo {author} {\bibfnamefont {Jianchong}\
			\bibnamefont {Xing}}, \bibinfo {author} {\bibfnamefont {Wenkai}\ \bibnamefont
			{Bai}}, \bibinfo {author} {\bibfnamefont {Bo}~\bibnamefont {Xiong}}, \bibinfo
		{author} {\bibfnamefont {Jun-Hui}\ \bibnamefont {Zheng}}, \ and\ \bibinfo
		{author} {\bibfnamefont {Tao}\ \bibnamefont {Yang}},\ }\bibfield  {title}
	{\enquote {\bibinfo {title} {Structure and dynamics of binary
				{B}ose–{E}instein condensates with vortex phase imprinting},}\ }\href
	{\doibase 10.1007/s11467-023-1316-0} {\bibfield  {journal} {\bibinfo
			{journal} {Frontiers of Physics}\ }\textbf {\bibinfo {volume} {18}},\
		\bibinfo {pages} {62302} (\bibinfo {year} {2023})}\BibitemShut {NoStop}%
	\bibitem [{\citenamefont {Bai}\ \emph {et~al.}(2021)\citenamefont {Bai},
		\citenamefont {Xing}, \citenamefont {Yang}, \citenamefont {Yang},\ and\
		\citenamefont {Liu}}]{RP.22.103828}%
	\BibitemOpen
	\bibfield  {author} {\bibinfo {author} {\bibfnamefont {Wen-Kai}\ \bibnamefont
			{Bai}}, \bibinfo {author} {\bibfnamefont {Jian-Chong}\ \bibnamefont {Xing}},
		\bibinfo {author} {\bibfnamefont {Tao}\ \bibnamefont {Yang}}, \bibinfo
		{author} {\bibfnamefont {Wen-Li}\ \bibnamefont {Yang}}, \ and\ \bibinfo
		{author} {\bibfnamefont {Wu-Ming}\ \bibnamefont {Liu}},\ }\bibfield  {title}
	{\enquote {\bibinfo {title} {Nonlinear dynamics of a {Bose-Einstein}
				condensate excited by a vortex ring phase imprinting},}\ }\href
	{https://www.sciencedirect.com/science/article/pii/S2211379721000152}
	{\bibfield  {journal} {\bibinfo  {journal} {Results in Physics}\ }\textbf
		{\bibinfo {volume} {22}},\ \bibinfo {pages} {103828} (\bibinfo {year}
		{2021})}\BibitemShut {NoStop}%
	\bibitem [{\citenamefont {Frisch}\ \emph {et~al.}(1992)\citenamefont {Frisch},
		\citenamefont {Pomeau},\ and\ \citenamefont {Rica}}]{PhysRevLett.69.1644}%
	\BibitemOpen
	\bibfield  {author} {\bibinfo {author} {\bibfnamefont {T.}~\bibnamefont
			{Frisch}}, \bibinfo {author} {\bibfnamefont {Y.}~\bibnamefont {Pomeau}}, \
		and\ \bibinfo {author} {\bibfnamefont {S.}~\bibnamefont {Rica}},\ }\bibfield
	{title} {\enquote {\bibinfo {title} {Transition to dissipation in a model of
				superflow},}\ }\href {\doibase 10.1103/PhysRevLett.69.1644} {\bibfield
		{journal} {\bibinfo  {journal} {Phys. Rev. Lett.}\ }\textbf {\bibinfo
			{volume} {69}},\ \bibinfo {pages} {1644--1647} (\bibinfo {year}
		{1992})}\BibitemShut {NoStop}%
	\bibitem [{\citenamefont {Raman}\ \emph {et~al.}(1999)\citenamefont {Raman},
		\citenamefont {K\"ohl}, \citenamefont {Onofrio}, \citenamefont {Durfee},
		\citenamefont {Kuklewicz}, \citenamefont {Hadzibabic},\ and\ \citenamefont
		{Ketterle}}]{PhysRevLett.83.2502}%
	\BibitemOpen
	\bibfield  {author} {\bibinfo {author} {\bibfnamefont {C.}~\bibnamefont
			{Raman}}, \bibinfo {author} {\bibfnamefont {M.}~\bibnamefont {K\"ohl}},
		\bibinfo {author} {\bibfnamefont {R.}~\bibnamefont {Onofrio}}, \bibinfo
		{author} {\bibfnamefont {D.~S.}\ \bibnamefont {Durfee}}, \bibinfo {author}
		{\bibfnamefont {C.~E.}\ \bibnamefont {Kuklewicz}}, \bibinfo {author}
		{\bibfnamefont {Z.}~\bibnamefont {Hadzibabic}}, \ and\ \bibinfo {author}
		{\bibfnamefont {W.}~\bibnamefont {Ketterle}},\ }\bibfield  {title} {\enquote
		{\bibinfo {title} {Evidence for a critical velocity in a bose-einstein
				condensed gas},}\ }\href {\doibase 10.1103/PhysRevLett.83.2502} {\bibfield
		{journal} {\bibinfo  {journal} {Phys. Rev. Lett.}\ }\textbf {\bibinfo
			{volume} {83}},\ \bibinfo {pages} {2502--2505} (\bibinfo {year}
		{1999})}\BibitemShut {NoStop}%
	\bibitem [{\citenamefont {Feynman}(1955)}]{Feynman}%
	\BibitemOpen
	\bibfield  {author} {\bibinfo {author} {\bibfnamefont {R.~P.}\ \bibnamefont
			{Feynman}},\ }\href@noop {} {\emph {\bibinfo {title} {Progress in Low
				Temperature Physics}}},\ edited by\ \bibinfo {editor} {\bibfnamefont {C.~J.}\
		\bibnamefont {Gorter}}\ (\bibinfo  {publisher} {North-Holland, Amsterdam,}\
	\bibinfo {year} {1955})\ p.\bibinfo {pages} {17}\BibitemShut {NoStop}%
\end{thebibliography}

%

\end{document}